\documentclass[journal=jacsat,manuscript=article]{achemso}

\usepackage[dvipsnames,x11names]{xcolor}
\usepackage{chemformula} % Formula subscripts using \ch{}
\usepackage[T1]{fontenc} % Use modern font encodings
\usepackage{bm} % bold math
\usepackage{graphicx}
\graphicspath{{figures/}}
\usepackage{caption}
\usepackage{subcaption}
\usepackage{cleveref}

\usepackage{tabularx}
\usepackage{longtable}
\usepackage{tikz}
\usepackage{multirow}
\usepackage{soul}
\usepackage[normalem]{ulem}
\usepackage[version=3]{mhchem} % Formula subscripts using \ce{}

\author{Merin Joseph}
\email{merin.joseph@nbi.ku.dk}
\affiliation{Department of Chemistry, Technical University of Denmark, Kgs. Lyngby, Denmark}
\alsoaffiliation{Niels Bohr Institute, University of Copenhagen, Denmark}
\author{Daniel J. Read}
\email{D.J.Read@leeds.ac.uk}
\affiliation{ School of Mathematics, University of Leeds, Leeds LS2 9JT, UK}
\author{Alastair M. Rucklidge}
\email{A.M.Rucklidge@leeds.ac.uk}
\affiliation{ School of Mathematics, University of Leeds, Leeds LS2 9JT, UK}

\newcommand{\revision}[1]{\textcolor{black}{{#1}}}
\newcommand{\revisioneq}{\color{black}}

\title[Flexible implementation of SST for ABC star terpolymers]
  {A flexible implementation of strong segregation theory for two dimensional ABC star terpolymer morphologies}

\keywords{Polymer phase separation, ABC terpolymers, Strong Segregation Theory}

\begin{document}

%%%%%%%%%%%%%%%%%%%%%%%%%%%%%%%%%%%%%%%%%%%%%%%%%%%%%%%%%%%%%%%%%%%%%
%% The "tocentry" environment can be used to create an entry for the
%% graphical table of contents. It is given here as some journals
%% require that it is printed as part of the abstract page. It will
%% be automatically moved as appropriate.
%%%%%%%%%%%%%%%%%%%%%%%%%%%%%%%%%%%%%%%%%%%%%%%%%%%%%%%%%%%%%%%%%%%%%
\begin{tocentry}
\includegraphics[scale=1]{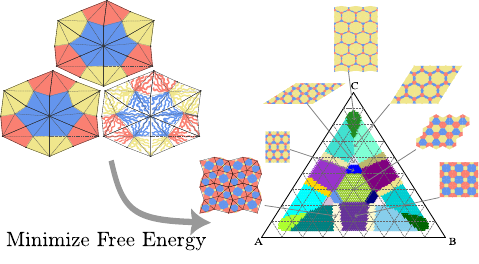}
\end{tocentry}

%%%%%%%%%%%%%%%%%%%%%%%%%%%%%%%%%%%%%%%%%%%%%%%%%%%%%%%%%%%%%%%%%%%%%
%% The abstract environment will automatically gobble the contents
%% if an abstract is not used by the target journal.
%%%%%%%%%%%%%%%%%%%%%%%%%%%%%%%%%%%%%%%%%%%%%%%%%%%%%%%%%%%%%%%%%%%%%
\begin{abstract}
We present a novel computational implementation of strong segregation theory, developed specifically for calculations of phase separated ABC star terpolymers. 
The method allows calculation of free energies of common two-dimensional morphologies for these polymers and the efficient construction of phase diagrams.
The branch points of the ABC star terpolymers are localized in core regions, modeled as cylinders in three dimensions, and our framework is applicable to morphologies with single and multiple core types.
Our central idea is that all the structures we wish to model can be assembled from a flexible base motif, which we call Strongly Segregated Polygons.
This method is useful for exploring a wide range of complex morphologies, using a range of compositions and interaction strengths.
We focus on 2D morphologies of ABC star terpolymers, but our method could be extended into three dimensions and to other molecular architectures, and in principle to large, irregular quasiperiodic two-dimensional structures. 
\end{abstract}

%%%%%%%%%%%%%%%%%%%%%%%%%%%%%%%%%%%%%%%%%%%%%%%%%%%%%%%%%%%%%%%%%%%%%
%% Start the main part of the manuscript here.
%%%%%%%%%%%%%%%%%%%%%%%%%%%%%%%%%%%%%%%%%%%%%%%%%%%%%%%%%%%%%%%%%%%%%

\section{Introduction}
    
In this paper, we present a novel computational implementation of strong segregation theory, developed specifically for calculations of phase separated ABC star terpolymers. 
The method allows calculation of free energies of common morphologies for these polymers and the construction of phase diagrams, but (because of its relative simplicity and computational efficiency) is in principle extendable to large, irregular quasiperiodic two-dimensional (2D) structures. 
Here, we will present the details and development of the method and its application to simple 2D morphologies.

Strong segregation theory, first developed by Semenov~\cite{Semenov1985}, is one of several methods that have been applied successfully in the study of the rich variety of block copolymer morphologies. 
The theory is applicable when the repulsive interaction strength between monomers is large enough that blocks are stretched away from thin interfaces between regions of different monomer type. 
This gives rise to a balance between interfacial energy and stretching energy, which can be computed for various morphologies, allowing a phase diagram to be constructed. 
This analytical method was further extended to include bicontinuous structures~\cite{Olmsted1998, Likhtman2000AnPolymers, Grason2022, Grason2023}  for diblocks, polymer brushes~\cite{Milner1994ChainMicrophases} and star-copolymers/miktoarms~\cite{Grason2004InterfacesCopolymers}. 

When the interaction strength is lower, the stability of phase separated morphologies is classically studied using other methods: linear theory~\cite{Joseph2023}, weak segregation theory~\cite{Leibler1980} and self-consistent field theory~\cite{Matsen1996}.
In these methods,  monomer density fluctuations are taken into account in the free energy calculation.  
When the interaction strength is very low, the block copolymer melt is almost homogeneous, and weak segregation theory, first developed by Leibler~\cite{Leibler1980}, is often applied. 
In this limit, the phase separated structures are \revision{often} described in terms of sinusoidal modulations in composition, retaining the lowest order interactions between different wavenumbers. 
For increased interaction strengths,  self-consistent field theory (SCFT), \revision{first developed by Helfand et al.~\cite{helfand1976block} and later extended to diblocks by Matsen et al.,~\cite{matsen1994stable}} is used. 
Here, the monomer composition is treated as a spatially varying field and the configuration distribution of chains interacting within that field is determined. 
Yet, the monomer composition must be self-consistent with the chain configurations, so the solution is iterated until the composition field and chain configurations agree.
The SCFT framework is the main workhorse for the study of microphase separation owing to its accuracy and validity in the intermediate and strong segregation limits\cite{Arora2016BroadlyDiscovery,GrasonMay2005,Grason2003,Cody2024StableInteractions}: it essentially overlaps with and bridges between the strong and weak segregation limits. 
The open-access PSCF software~\cite{Arora2016BroadlyDiscovery} is widely used to implement  SCFT calculations for a variety of block copolymer systems.

All three nonlinear methods agree in the ordering of classical diblock morphologies (lamellae, rods, gyroid and spheres) in the simple diblock case \cite{Olmsted1998, Leibler1980, Matsen1996}. 
It is nevertheless computationally intensive to perform SCFT calculations, especially as the molecular complexity and size of the simulation box increases~\cite{Duan2018StabilityCopolymers, LiW2010}.
Under these circumstances, it remains useful to return to simpler theories, such as strong segregation theory, to screen a wide variety of compositions and interaction strengths before undertaking SCFT calculations. 
By developing a simple computational implementation of strong segregation theory, we are here aiming towards a useful tool that will allow a wide variety of morphologies to be investigated with relative computational ease.

\begin{figure}
\begin{center}
\includegraphics[width=1\linewidth]{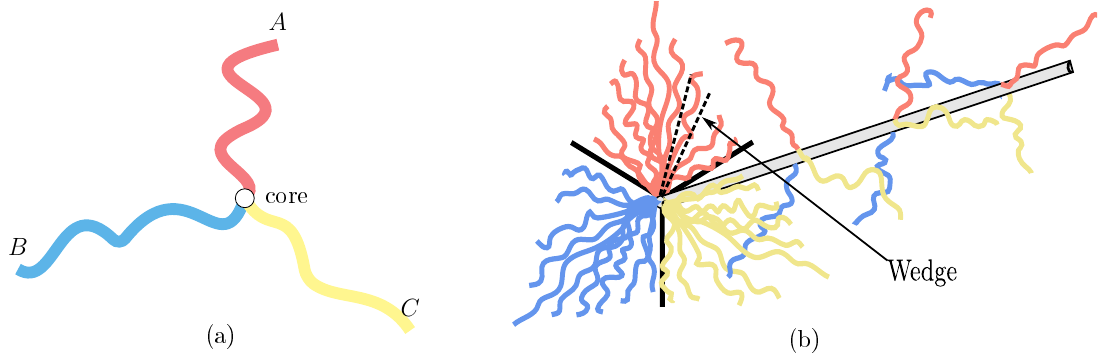}
		\end{center}
		  \caption[ABC star structure and wedge]{A graphical representation of (a)~an ABC star terpolymer and (b)~phase separation in ABC star terpolymer melts.
      The three branches (red, blue and yellow) join at their core, and in three dimensions, the different polymer types collect into three domains, with the cores aligned in a straight line, forming columnar structures in one dimension and ordered geometrical patterns in the other two.
    The cores form a structure that can be taken as a narrow cylinder (gray) to which the polymer branches are grafted.
    The radius of this cylinder is comparable to the length of a monomer unit.
    The interfaces between the three domains are indicated by thick black lines.
    The stretching free energy is calculated by slicing each domain into small wedges, indicated by dashed lines. }
  \label{fig:abcterpolymer}
		\end{figure}
        
ABC star terpolymers are synthesized by grafting a third block (of type C) to a diblock at the junction where the A and B blocks meet~\cite{Fujimoto1992PreparationBranches, Huckstiidt1996SynthesisCopolymer, Iatrou1992SynthesisTerpolymer, Sioula1997SynthesisMethacrylate, Lambert1998SynthesisCopolymers, Huckstadt2000SynthesisPoly2-vinylpyridine, Pennisi1988PreparationStars}. 
This results in a branched block copolymer, illustrated in \cref{fig:abcterpolymer}(a). 
If the degree of polymerization of all three blocks is sufficiently large and the chemistries sufficiently incompatible, then melts of such molecules tend to microphase separate into regions rich in each of the A, B or C blocks \cite{Okamoto1997MorphologyCopolymers, Huckstiidt1996SynthesisCopolymer, Sioula1998DirectStructure}. 
This results in a wide variety of phases that have been investigated experimentally, computationally and theoretically. 
We present a summary of experimental results in \cref{table:experimentmorphologies} and computational/theoretical results in \cref{table:computational studies}.

When A-, B- or C-rich domains are strongly segregated, molecules must be configured in such a way that the A, B and C blocks can extend into their respective domains. 
Hence, the junction point of the stars must be located at places where all three domain types meet. 
The meeting of three domains in 3D space will trace out a space curve, so we anticipate that junction points of stars are located at these space curves, as illustrated in \cref{fig:abcterpolymer}(b). 
We expect a large number of star junction points to be located along each of these curves and will henceforth refer to these curves as the ``core'' regions of the structures. 

In many (but not all) cases, structures formed are quasi-2D in nature, where a phase-separated pattern in 2D is extended, unchanged, into the third dimension. 
In such cases, the ``cores'' are points in 2D where three domain types meet, and form straight lines in the third dimension (see \cref{fig:abcterpolymer}(b)). 
These quasi-2D structures will be the main focus of this paper.
They can be depicted as 2D tilings of A-, B- and C-rich regions, with the areas of the different domains related to the degree of polymerization of each monomer type. 
Since these tilings must contain  the cores at the intersection of all three domains, each domain type (e.g., A) must be surrounded by alternating domains of the other two types (e.g., B and C) with the cores located at the points (in 2D) where all three domain types meet.

\begin{figure}
\begin{center}
\includegraphics[width=1\linewidth]{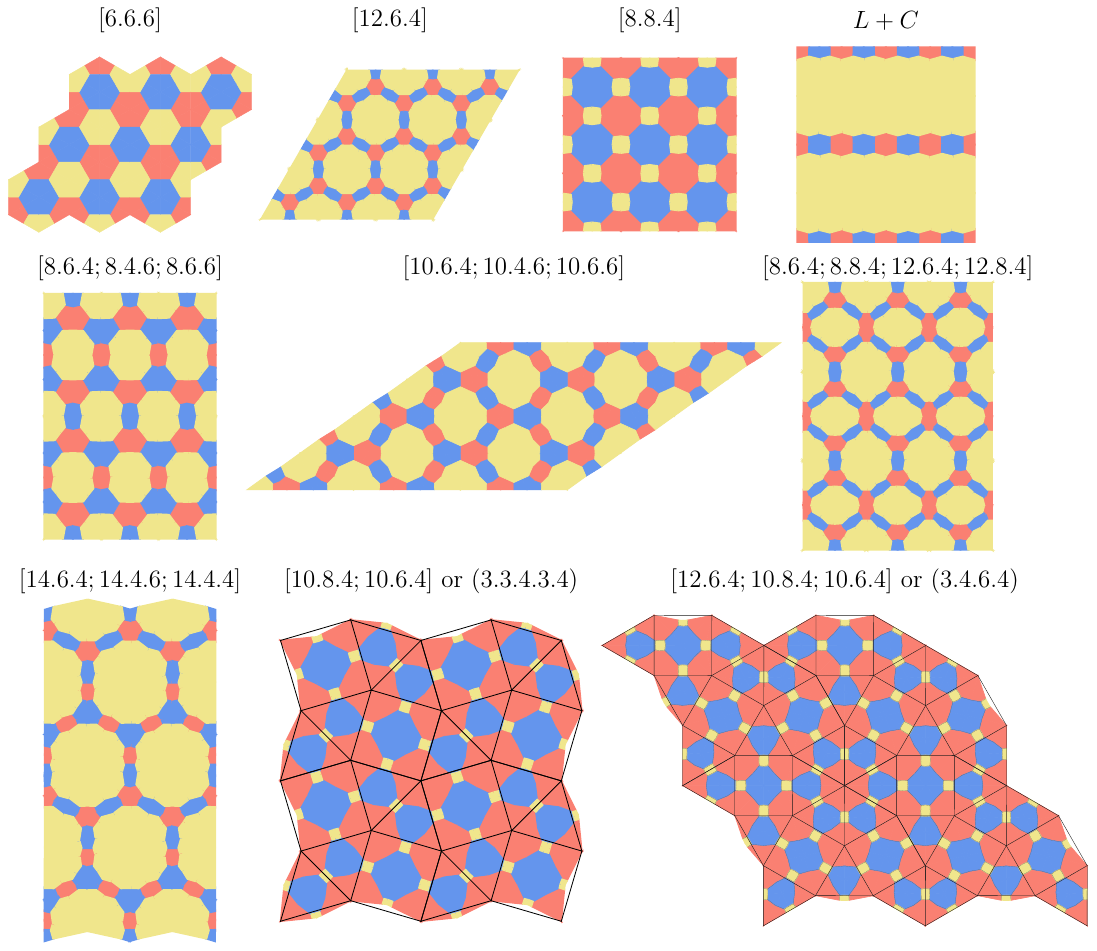}
\end{center}
\caption[Morphologies]{\revision{All candidate morphologies considered in this work are illustrated here.
The color scheme is A (red), B (blue) and C (yellow), and the areas of each domain are related to the lengths of the A, B and C branches.
The cores are at the junctions of the red, blue and yellow regions.
The morphologies are all periodic in space, and several repeats of the periodic morphologies are shown (four repeats in the last two examples).
The last two morphologies are decorated (in black) with square and triangular tiles, which assist in constructing the morphologies.}}
  \label{fig:morphologies_withcore}
		\end{figure}

Even with this restriction, a rich variety of phases is possible, and several of the commonly investigated periodic structures are illustrated in \revision{\cref{fig:morphologies_withcore}}.
The commonly used notation for most of these uses Schläfli symbols~\cite{Grunbaum1987}, which are triplets of three numbers $[x.y.z]$ indicating the structure.
These numbers refer to the number of neighbouring domains surrounding each domain type. 
The simplest structures are depicted by only one triplet of numbers, e.g., $[8.8.4]$ is a morphology with square symmetry in which two of the domain types (e.g., A and B) are surrounded each by eight neighbouring domains, whilst the third type (e.g.,~C) is surrounded only by four (see \revision{\cref{fig:morphologies_withcore}}). Morphologies $[6.6.6]$ and $[12.6.4]$ similarly require only a single triplet of numbers. 
These structures are simple in the sense that all cores in the structure are identical, being surrounded by equivalent arrangements of A, B and C domains. 
The ``L+C'' (lamellae + cylinder) phase depicted in \revision{\cref{fig:morphologies_withcore}} is also simple in the same sense (having only one core type) but is not usually denoted by the triplet notation since the lamellar domains are surrounded by an infinite number of neighbours.

More complicated structures are denoted by multiple triplets because there are multiple types of domain for each given monomer. 
As a result, these structures also contain multiple types of core that are not equivalent.  
One way of labelling these is to consider the three domains adjacent to each core, and to note the number of neighbours connected to each of those three neighbouring domains. 
For example, we denote a three-core structure by $[8.6.4; 8.4.6; 8.6.6]$; see \revision{\cref{fig:morphologies_withcore} and \cref{fig:candidate_morphologies}} for more detail.
This structure contains some cores adjacent to an 8-neighbour \revision{C}~domain, 6-neighbour \revision{B}~domain and 4-neighbour \revision{A}~domain; some cores adjacent to an 8-neighbour \revision{C}~domain, 4-neighbour \revision{B}~domain and 6-neighbour \revision{A}~domain; and some cores adjacent to an 8-neighbour \revision{C}~domain, 6-neighbour B~domain and 6-neighbour \revision{A} domain. 
\revision{Although in this case it is the C~domains that have 8~neighbours, the standard Schläfli notation is to list the highest number first, regardless of the polymer type.}
Even this notation is not exhaustive: there are actually two different types of $[8.6.6]$ core, not equivalent by symmetry, within this structure. 

Another important structure is the so-called ``$\Sigma$-phase'', also denoted by the Archimedian tiling notation $(3.3.4.3.4)$~\cite{Grunbaum1987}, which we label as $[10.8.4; 10.6.4]$, while noting that there are several types of each of the $[10.8.4]$ and $[10.6.4]$ cores not equivalent by symmetry. 
As we will see below, the fact that there are multiple core types within a single structure is an important consideration when minimizing the energy: different core types will typically contain a different number of star centres per unit length.

ABC star terpolymers have been synthesized from different monomers, and phase separated morphologies are observed in most examples.
We give a sample of these in \cref{table:experimentmorphologies}.
The majority of morphologies found in this wide range of experiments are single core.
Multi-core structures typically require blends of ABC star terpolymers with other components.
\Cref{table:computational studies} reports a sample of computational and theoretical studies of morphologies of ABC star terpolymers.
These studies report a variety of quasi-2D morphologies and some 3D morphologies, finding a wider range than identified in experiments.

\begin{table}[H]
\begin{tabular}{|p{4cm}|p{3.5cm}|p{8cm}|}
 \hline 
 Article  & Polymer types    &  Morphologies identified   \\
 \hline
 Hadjichristidis et al.\cite{Hadjichristidis1993MorphologyTerpolymers} & PS, PI, PB  &  Cylinders of PS and miscible matrix of PI and PB\\
\hline
Sioula et al.  \cite{Sioula1998,Sioula1998DirectStructure} &  PS, PI, PMMA &  Coaxial cylinders in 2D  hexagonal lattice   \\
\hline
Okamoto et al.\cite{Okamoto1997MorphologyCopolymers}  &PS, PDMS, PTBMA  & $[6,6,6]$\\
\hline
Hückstädt et al.\cite{Huckstadt2000SynthesisPoly2-vinylpyridine}  & PS, PB, PVP & $[8.8.4]$, $[12.6.4]$, Lamellar + Coaxial cylinder\\
\hline
Yamauchi et al.\cite{Yamauchi2003MicrodomainTomography}  &PS, PI, PDMS  & $[8.8.4]$\\
\hline
Takano et al. \cite{Takano2004,Takano2005}  & Blends of PI, PS, PVP & $[6.6.6]$, $[8.8.4]$, $[12.6.4]$, $[10.8.4; 10.6.4]$\\
\hline
Hayashida et al. \cite{Hayashida2006,Hayashida2007a}& PI, PS, PVP blends & Dodecagonal quasicrystals and $[10.8.4; 10.6.4]$\\
\hline
Aissou et al. \cite{Aissou2013OrderedTerpolymer} and  Nunns et al. \cite{Nunns2013SynthesisMetalloblock}  &PI, PS, PISF  & $[8.8.4]$, $[6.6.6]$\\
\hline
Choi et al. \cite{KyoonChoi2014ThinTerpolymer}   &PI, PS, PISF  & L+C\\
\hline
Chernyy et al.\cite{Chernyy2018Bulk}   &PDMS, PI, PMMA &  $[6.6.6]$, $[8.8.4]$, L+C and columnar discs \\
\hline
Ariaee et al.\cite{Ariaee2023}   &PI, PS, PMMA &  L+C, Coaxial cylinders in 2D  hexagonal lattice, Coaxial cylinders in 2D  square lattice  \\
\hline
\end{tabular} 
\caption{A summary of morphologies reported experimentally in ABC star terpolymer melts.}
 % \mj{polystyrene (PS), polyisoprene (PI), and polymethyl methacrylate (PMMA)}}
\label{table:experimentmorphologies}
\end{table}

 % \begin{table}[H]

\begin{longtable}{|p{3cm}|p{3cm}|p{9.5cm}|}
\hline
 Article & Computational framework   &  Morphologies identified   \\
 \hline
Dotera et al. \cite{Dotera1996TheCopolymers} & Monte Carlo (Diagonal Bond Method) &   $[6.6.6]$\\
\hline
Bohbot et al.~\cite{BohbotRaviv2000a} & SCFT & $[6.6.6]$, $[8.8.4]$\\
\hline
Gemma et al. \cite{Gemma2002} & Monte Carlo (Diagonal Bond Method) & Lamella + sphere (L+S), $[8.8.4]$, $[6.6.6]$, $[8.6.4; 8.4.6; 8.6.6]$,
$[10.6.4; 10.4.6; 10.6.6]$, $[12.6.4]$, perforated layer (PL), L+C, columnar piled disk (CPD), and lamella-in-sphere (L-in-S)\\
\hline
He et al. \cite{He2002,He2003LocalizationsTerpolymers}  & Dynamic Density Functional Theory  & $[6.6.6]$, $[8.4.4]$, $[12.6.4]$, coaxial cylinders, lamellae \\
\hline
Ueda et al.\cite{Ueda2007}  & Monte Carlo (Diagonal Bond Method) & $[6.6.6]$, $[8.4.4]$, $[12.6.4]$, coaxial cylinders, L+S, L+C, $[10.8.4; 10.6.4]$ \\
\hline
Birshtein et al. \cite{Birshtein2004TheoryStar-terpolymer} & SST & Lamellae  \\
\hline
Dotera et al.\cite{Dotera2006} &  Monte Carlo (Diagonal Bond Method) &  12-fold quasicrystals\\
\hline
Huang et al. \cite{Huang2007MicrophaseDynamics} & Dissipative Particle Dynamics (DPD) & $[6.6.6]$, $[8.4.4]$, $[12.6.4]$, coaxial cylinders, L+S, L+C, $[10.8.4; 10.6.4]$, disordered networks\\
\hline
Li et al \cite{LiW2010} & SCFT &  L+S, $[8.8.4]$, $[6.6.6]$, $[8.6.4; 8.4.6; 8.6.6]$,
$[10.6.4; 10.4.6; 10.6.6]$, $[12.6.4]$, $[10.8.4; 10.6.4]$, PL, L+C, CPD \\
\hline
Zhang et al.\cite{Zhang2010}&  SCFT & $[8.8.4]$, $[6.6.6]$, $[8.6.4;8.4.6; 8.6.6]$,
$[10.6.4; 10.4.6; 10.6.6]$, $[12.6.4]$, $[8.6.4; 8.8.4; 12.6.4; 12.8.4]$, L+C, lamellae\\
\hline
Xu et al. \cite{Xu2013ACopolymers}&  SCFT & $[8.8.4]$, $[6.6.6]$, $[8.6.4;8.4.6; 8.6.6]$,
$[10.6.4; 10.4.6; 10.6.6]$, $[12.6.4]$, $[8.6.4; 8.8.4; 12.6.4; 12.8.4]$, $[10.8.4; 10.6.4]$, L+C, lamellae and disordered networks \\
\hline
Kirkensgaard et al.\cite{Kirkensgaard2014}&  Dissipative Particle Dynamics (DPD) & $[8.8.4]$, $[6.6.6]$, $[8.6.4; 8.4.6; 8.6.6]$,
$[10.6.4; 10.4.6; 10.6.6]$, $[12.6.4]$, $[12.6.6; 12.6.4; 12.4.4]$, $[10.8.4; 10.6.4]$, L+C, lamellae\\
\hline
Jiang et al. \cite{Jiang2015}  & SCFT & $[8.8.4]$, $[6.6.6]$, $[8.6.4; 8.4.6; 8.6.6]$,
$[10.6.4; 10.4.6; 10.6.6]$, $[12.6.4]$, $[8.6.4; 8.8.4; 12.6.4; 12.8.4]$, $[10.8.4; 10.6.4]$, L+C, lamellae, disordered networks\\
\hline
Hawthorne et al. \cite{Cody2024StableInteractions} &  SCFT & $[8.8.4]$, $[6.6.6]$, $[8.6.4; 8.4.6; 8.6.6]$,
$[10.6.4; 10.4.6; 10.6.6]$, $[12.6.4]$, $[8.6.4; 8.8.4; 12.6.4; 12.8.4]$, $[10.8.4; 10.6.4]$, L+C, lamellae, cylinders\\
\hline
% \end{tabular} 
\caption{A summary of morphologies in ABC star terpolymers reported using different computational methods.}
\label{table:computational studies}
\end{longtable}

In this paper, we focus specifically on the 2D morphologies of ABC star terpolymer melts, and particularly on developing a computational framework for implementing a strong segregation theory for these morphologies.  
Previously, Gemma {\textit et al.}\cite{Gemma2002} constructed a strong segregation theory (SST) for some specific structures, those with a single core type. 
In our work, we aim to generalise this, creating a framework that is applicable to morphologies with multiple core types, more disordered patterns, or quasicrystalline approximants. 
Key to all these structures is the existence of cores, as described above, with the star arms strongly stretched away from the cores into their respective A-, B-, or C-rich domains. 

Our central idea is that all the structures we wish to model can be assembled from a flexible base motif structure, illustrated in \cref{ch1fig:sspframework}(a) below. 
We call these motif structures ``strongly segregated polygons'' (SSPs).
Each SSP contains a single core, surrounded by A, B and C domains, and we represent its shape by a polygon. 
The minimal number of edges is six, giving a balance between flexibility and complexity. 
Each six-sided SSP requires six outer nodes to specify its shape, so that it tessellates with the other polygons to form the overall morphology. 
In our method, we tessellate a periodic domain with SSPs, compute the free energy of the configuration, and minimize this free energy by adjusting the positions of the boundary points of the SSPs. 
In structures with only one core type, all these SSPs are generally identical (up to symmetry) in a minimal energy structure. 
However, in more complicated structures with multiple core types, there will naturally be variation between the SSPs, both in shape and area. 

We anticipate that this method will be useful for exploring a wide range of complex morphologies, using a range of compositions and interaction strengths.
While SST is only valid in the limit of high interaction strengths, the distribution of domains of monomer types can be used as initial conditions for more accurate calculations using (for example) \hbox{SCFT}.
We focus on 2D morphologies of ABC star terpolymers, but our SSP method could be extended into three dimensions and to other molecular architectures.
\revision{As an example of this for AB block copolymers, Reddy, Dimitriyev and Grason \cite{Grason2022, Grason2023} constructed a strong segregation theory for bicontinuous phases such as double-gyroid, in which the interconnected structure is composed of ``meso-atoms''. To compute the chain conformations and free energy within each meso-atom, they generalised the idea of ``wedges'' used previously by Olmsted and Milner \cite{Olmsted1998}. Each wedge comprises an extended volume terminating in two triangular faces, with an internal triangular facet separating the A and B rich domains, such that the A and B blocks stretch away from the internal facet. Hence, the wedge is the tessellating base motif structure for an AB copolymer, analogous to our SSPs.}

The rest of this paper is devoted to: establishing the free energy for a single SSP; demonstrating that candidate morphologies can be constructed from an assembly of SSPs; minimizing the free energy of each morphology by moving the nodes of each SSP; and creating phase diagrams for the ABC star terpolymer melt with several choices of parameter values.

\section{Methodology}
In the strong segregation limit, the free energy of a branched block copolymer system depends only on the geometry of the domains of the different blocks and the repulsion between monomer units of different types.
The calculations needed for strong segregation theory are relatively easy (compared to other mean field approximations) and can be carried out for a wide variety of morphologies\cite{Milner}.

  \subsection{Free energy calculation}
  
In this work, we focus on ABC star terpolymers as illustrated in \cref{fig:abcterpolymer}(a), with the A, B and C branches joining at a core.
We assume that the cores of the polymer molecules align in straight lines, with domains of A, B and C surrounding each line, giving a two dimensional pattern when viewed from the third direction.
We consider a line of cores as a cylinder in three dimensions, of radius~$R_{core}$ \revision{expected to be} comparable to the monomer size~$b$, with the three polymer types grafted to the cylinder, as shown in \cref{fig:abcterpolymer}(b). \revision{We separate the free energy calculation into three parts: the stretching energy for chains outside the core region; consideration of the core itself; and interfacial energy due to surface tension between the A, B and C domains.}

In order to evaluate the stretching energy of the polymer branches, we divide the three regions into wedges, as shown in \cref{fig:abcterpolymer}(b), with wedges containing polymers grafted onto the inner cylinder.
The stretching energy for polymer brushes grafted onto a convex surface of radius of curvature~$R_{core}$ is reported by Ball et al.\citep{Ball1991}
In a wedge of height~$h$ containing monomers of type~$I$ in blocks of length~$N_I$, the stretching free energy per chain in units of~$k_BT$ is 
 \begin{equation} \label{eq:Ball_stretching}
   \frac{3}{4}\frac{h^2}{N_I b^{2}}\log{\frac{h^2}{R_{core}^2}},
 \end{equation}
where we have taken the leading term in the limit of small~$R_{core}$.\cite{Ball1991,Joseph2023searching}
For convenience, from this point onwards, we express all lengths in units of $R=\sqrt{Nb^2}$, where $N=N_A+N_B+N_C$ is the total number of monomers in a star, and we define $\phi_I=N_I/N$.
Then the stretching energy per chain in a wedge becomes
\begin{equation}\label{ch3eq:Stretching_with_convex}
   f_{chain}(H,\phi_I) = \frac{3}{4\phi_I}H^2\log(cH^2) = \frac{3}{4\phi_I}F_{chain}(H),
\end{equation}
where $H=h/R$, $c=R^2/R_{core}^2$ and $F_{chain}(H)=H^2\log(cH^2)$.
The stretching free energy of the whole morphology can then be found by summing over all the wedges with different heights and different monomer types.

\begin{figure}
\begin{center}
\includegraphics[width=1\linewidth]{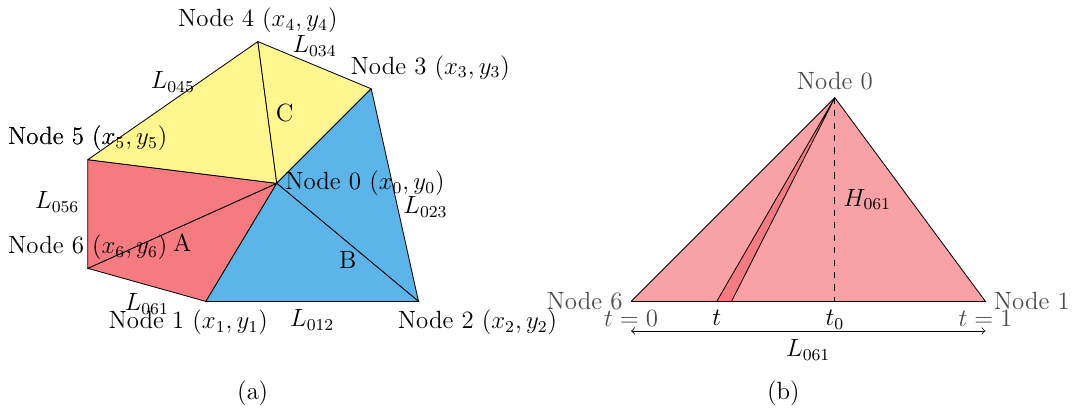}
\end{center}
\caption[SSP framework]{Geometrical structure of a Strongly Segregated Polygon (SSP). In~(a), the six-sided SSP has one node (Node~0) in the center and six nodes around the edges.
The three colors indicate the regions of three types of polymer.
In~(b), we show triangle~061 of the SSP, with a coordinate~$t$ that runs from $0$ to~$1$ along the bottom of the triangle.
The dimensionless height of the triangle is $H_{061}$ and the length of its base is~$L_{061}$.
A wedge (at position~$t$) within this triangle is illustrated.}
  \label{ch1fig:sspframework}
		\end{figure}

In order to evaluate this sum, we will describe the two-dimensional morphologies formed by ABC terpolymers in terms of polygons, with a core cylinder at the center of each polygon.
One such Strongly Segregated Polygon (SSP) is shown in \cref{ch1fig:sspframework}(a): it is made up three domains, one for each polymer type, with the domains divided in two, resulting in six triangles.
We assume that all polymers within each domain are attached to the core at Node~0 and stretched towards the edge of the domain, and that the number of polymers in a wedge is proportional to the area of the wedge.
The stretching free energy per chain in each triangle is then the integral over the wedges making up that triangle (\cref{ch1fig:sspframework}(b)). 
The interfacial energy comes from the boundaries between the domains of the three different polymer types within each~\hbox{SSP}.
There is no interfacial energy contribution between two SSPs because we assemble these so that they join at regions containing the same monomer type.

The \revision{stretching free energy, interfacial energy and core energy} of one SSP depend on the location of all its vertices, the monomer compositions ($\phi_A$, $\phi_B$ and~$\phi_C$) and the interaction strengths (expressed in terms of the Flory interaction parameters as $N\chi_{AB}$, $N\chi_{BC}$ and~$N\chi_{AC}$).
Given the positions of Nodes~1 to~6, the position of Node~0 at the center is determined uniquely from the monomer compositions, with the constraint that Node~0 should lie within the polygon (see \cref{ch3eq:x0y0} in the Supporting Information).

To determine the stretching energy per chain for a single triangle, we divide the triangle into wedges (see \cref{ch1fig:sspframework}(b)).
We give the outline of the calculation here, with further details in the Supporting Information.
We take the $061$ triangle as an example, containing polymer~\hbox{A}.
This triangle has base length $L_{061}$ and height $H_{061}$, measured in units of~$R=\sqrt{Nb^2}$.
The triangle has a coordinate~$t$ that runs from $0$ to~$1$ along its base, and $t_0$ gives the location of the perpendicular projection of node~$0$ onto the base.
The height of a wedge (in units of~$R$) from node~$0$ to the point parameterized by~$t$ is $H(t)=\sqrt{(t-t_0)^2 L_{061}^2+H_{061}^2}$.
The stretching energy per chain for the whole triangle is then
 \begin{equation}\label{stretching_free_triangle061}
 f_{061}=\frac{A_{061}}{\phi_{A}A_T} 
           \int_{0}^{1}
           f_{chain}\left(H(t), \phi_A\right)dt
        =\frac{3A_{061}}{4\phi_{A}^2A_T} 
           \int_{0}^{1}
           F_{chain}\left(H(t)\right)dt,
 \end{equation}
where $A_{061}$ is the area of the triangle and $A_T$ is the area of whole~\hbox{SSP}, in units of~$R^2$.
The function~$F_{chain}(H)=H^2\log(cH^2)$ and $c=R^2/R_{core}^2$ from above.

The second integral in~\cref{stretching_free_triangle061} can be written explicitly as
\begin{equation}\label{eq:Ifor061}
	\begin{aligned}
        I_{061}(X,Y,t_0)={}&\int_{0}^{1}\left((t-t_0)^2 X+Y\right)\log\left(c((t-t_0)^2 X+Y)\right)dt \\
    {}={}&\log(c(X(1-t_0)^2+Y))\left(\frac{X(1-t_0)^3}{3}+Y(1-t_0)\right)+{}\\
    &\log(c(Xt_0^2+Y))\left(\frac{Xt_0^3}{3}+Yt_0\right)-\frac{2X}{9}\left((1-t_0)^3+t_0^{3}\right)-\frac{4Y}{3}+{}\\
	&\frac{4Y}{3}\sqrt{\frac{Y}{X}}\left(\arctan\sqrt{\frac{X}{Y}}(1-t_0)+\arctan\sqrt{\frac{X}{Y}}t_0\right),
	\end{aligned}
\end{equation}
where $X=L_{061}^2$, $Y=H_{061}^2$ and $c=R^2/R_{core}^2$.
The stretching free energies for the remaining five triangles are calculated in a similar manner: triangle $0ij$, between nodes $0$, $i$ and~$j$, has base length $L_{0ij}$ and height $H_{0ij}$, so the stretching free energy is $f_{0ij}$.
The stretching free energy per chain of the entire SSP, in units of~$k_BT$, is then
\begin{equation}\label{ch3eq:SSPstrechfree}
	\begin{split}
		f_{str}=\frac{3}{4}\frac{1}{A_T}&\left(\frac{A_{061}I_{061}+A_{056}I_{056}}{ \phi_{A}^2}+
		\frac{A_{023}I_{023}+A_{012}I_{012}}{ \phi_{B}^2}+
		\frac{A_{045}I_{045}+A_{034}I_{034}}{ \phi_{C}^2}\right),
	\end{split}
\end{equation}
where, as in \cref{ch1fig:sspframework}(a), triangles 061 and 056 contain polymer~A, triangles 023 and 012 contain polymer~B, and triangles 045 and 034 contain polymer~\hbox{C}.

\revision{So far we have not considered the free energy of the core region itself, nor specified how large its radius $R_{core}$ might be.  Since we expect the core region to be of order of the monomer dimension, the exact details must depend on the specific chemistry, both of the monomer type in each of the three arms, and of the chemical unit used to form the branch point itself. Hence, it is not possible to develop a ``universal'' theory.  However, we can make reasonable assumptions to arrive at a plausible first-order description of the core. We assume that the chains in the core region are sufficiently closely packed together so that the arms are forced to exit the core region as quickly as possible, i.e., they are strongly stretched away from the core at the monomer scale. Thus, if the core region contains $N_{core} \ll N$ monomers per chain, we expect $R_{core} \approx N_{core}b/3$ for a three arm star. By equating the volume $\pi R_{core}^2 d$ of a core section of length $d$ with the volume occupied by the $N_{core}$ monomers per chain (see the Supplemental Information) we arrive at the estimates:}
\begin{align*}
  \revisioneq N_{core} & \revisioneq \approx \frac{9A_T}{\pi}, \\
   \revisioneq R_{core} &  \revisioneq \approx \frac{3A_T b}{\pi},
\end{align*}
\revision{where, as above, $A_T$ is the area of whole~\hbox{SSP}, in units of~$R^2$. As $A_T$ increases, the number of chains per unit length inside the core region also increases, so that the radius of the core region (in which chains are stretched to monomer level by chain packing) must increase. Hence, we find:}
\begin{equation}\label{eq:expression_for_c}
   \revisioneq c = \frac{R^2}{R_{core}^2} \approx \frac{\pi^2 N}{9A_{T}^2},
\end{equation}
\revision{i.e., the value of $c$ used in \cref{eq:Ifor061} above is not constant but varies with the SSP area, and depends on the degree of polymerization~$N$ as well as the SSP geometry via the normalised area $A_T$.}

\revision{This dependence on $N$ should be briefly commented on. In most theories of polymer phase separation, the dependence on $N$ arises in combination with the interaction parameters $\chi$, so that the product $N \chi$ is the relevant parameter. In our theory, $N \chi$ remains the leading order parameter, emerging as described below in the interfacial energy. The further dependence on $N$ (independent of the interaction parameters $\chi$) arises because of the explicit dependence on $R_{core}$ in \cref{eq:Ball_stretching}, giving a logarithmic dependence on $N$ in the stretching energy via the parameter $c$ in \cref{eq:Ifor061}. As a result, varying $N$ has a weak effect on the balance between stretching and interfacial energies, giving rise to small changes in the phase diagram, as we will describe below.}

\revision{We expect the free energy for the core region to result from two main contributions: a stretching energy of order $k_B T$ per monomer from stretching the chains at the monomer scale, and a mixing energy arising from forcing the A, B and C monomers into close proximity in the core region. Thus, we estimate the core energy per chain, in units of $k_B T$, to be proportional to $N_{core}=9A_T/\pi$ with form:}
\begin{equation}\label{eq:free_energy_core}
  \revisioneq  f_{core} = \frac{9A_T}{\pi} \left( s_{core} + \frac{\chi_{AB}+\chi_{BC}+\chi_{AC}}{9} \right) 
\end{equation}
\revision{where $s_{core}$ is an order one parameter related to the change in free energy, in units of $k_B T$, for fully orienting a single monomer in the direction away from the core. To obtain the second ``mixing'' term inside the brackets, we assume the core to be comprised of equal fractions of all three monomer species.  Since $A_T \sim H^2$, where $H$ is the (non-dimensional) degree of stretching of the chains in the SSP, this core energy acts in a largely similar way to the stretching energy. Since $s_{core}$ is of order one, and since experimentally reported Flory interaction parameters $(\chi)$ in the ABC terpolymer systems listed in \cref{table:experimentmorphologies} have values less than $0.1$, we take}
\begin{equation}
\revisioneq f_{core} = \frac{9}{\pi}A_T
\end{equation}
\revision{in all calculations reported in this paper.}

\revision{Although the above description of the core region is necessarily approximate, we believe none of our results are sensitive to the approximations used. Indeed, setting $c$ to be a reasonable constant, and setting $f_{core}=0$, results in the same geometry of phase diagrams and same form of free-energy minimized morphologies as we present below. This is because any quantitive changes are similar across all morphologies.}

The interfacial energies at the three interfaces, AB, BC and~AC, contribute to the interaction free energy.
The Flory interaction parameter $\chi$ is proportional to the square of the surface tension~$\gamma$ at the interfacial plane \cite{Milner1988}.
The interfacial energy per chain $f_{int}$ in units of $k_BT$ is then
\begin{equation}\label{eq:interfacial_energy_per_chain_polygon}
 f_{int}=\frac{\gamma_{AB}L_{01}+\gamma_{BC} 
 L_{03}+\gamma_{AC} L_{05}}{A_T},
 \end{equation}
where the interfaces have lengths (in units of~$R$) $L_{01}$ between A and~B, $L_{03}$ between B and~C, and $L_{05}$ between A and~C (see \cref{ch1fig:sspframework}), and the three scaled surfaces tensions are $\gamma_{AB}$, $\gamma_{BC}$ and~$\gamma_{AC}$.
These are related to the Flory interaction parameters by
\begin{equation}
    \gamma_{IJ} = \sqrt{\frac{N\chi_{IJ}}{6}}.
\end{equation}
See the Supporting Information for details.

The total free energy per chain for an \hbox{SSP} $f_c$ in units of $k_BT$ is
\begin{equation}\label{ch3eq:freeenergyf_cSSP}
	 \revisioneq f_c=f_{int}+f_{core}+f_{str}.
\end{equation}
This free energy for a single SSP is thus a function of monomer composition, interaction strengths and node coordinates.
A given morphology is initially set up as a combination of SSPs, and then the node coordinates can be adjusted to find a configuration that minimizes the total free energy per chain for the chosen monomer compositions and interaction strengths.

The discussion so far has been for six-sided SSPs, which we use for the majority of our calculations. However, lamellar phases (in particular, the L+C phase) cannot be represented with six-sided SSPs, so we extend the SSP method to allow for eight-sided SSPs, in which one of the domains is represented by four triangles, with two other domains represented with two triangles as before. The calculation of the stretching free energy within each triangle remains as before.

\begin{figure}
\begin{center}
\includegraphics[width=1\linewidth]{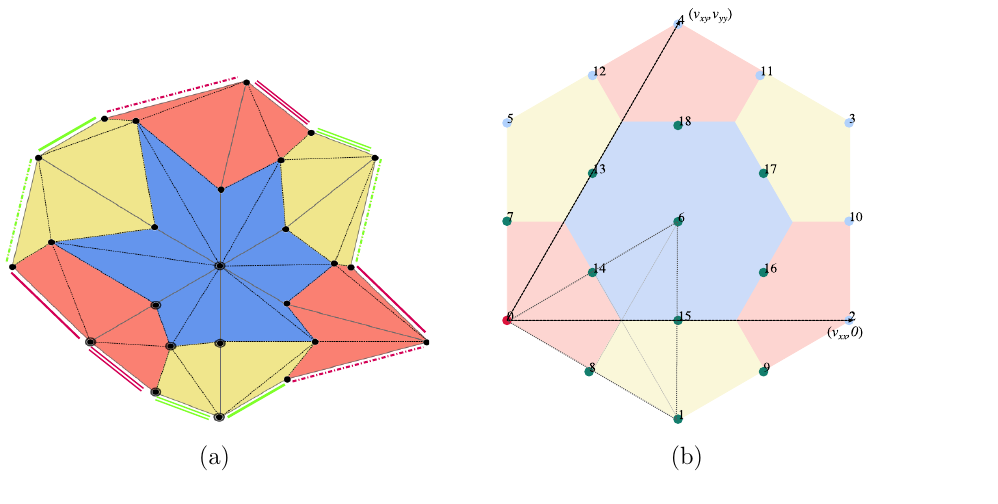}
		\end{center}
\caption[SSP framework]{Geometrical structures of SSPs.
In~(a) we demonstrate how to place six SSPs to form a periodic patch of the $[6.6.6]$ morphology.
The black dots indicate the nodes of the SSPs, and one SSP is marked out with larger dots.
Colored lines along edges indicate matching rules which makes the morphology periodic.
In~(a), the same type of lines indicates joining of edges to make the periodic patch.
In~(b) we give an illustration of the SSP optimisation framework. 
The pink dot is the reference node, green dots are the free nodes and light blue dots are the periodic nodes.
The black arrows indicates the periodicity, $[v_{xx},v_{xy},v_{yy}]$ of the repeating patch.
One \hbox{SSP} is indicated by dashed lines.}
  \label{fig:sspimplementation}
		\end{figure}

\subsection{Implementing the SSP method}

A required morphology is constructed using these SSPs by joining several of them together.
In joining two SSPs, a region containing polymer type~A in one SSP will join to a region containing type~A in the other; similarly B and~\hbox{C}.
In addition, nodes are shared between two or more adjacent~SSPs, and a change in location of a node will affect all SSPs that it belongs to.
We are always concerned with morphologies that are periodic, so the matching of polymers and the identification of nodes carries across the sides of the periodic domain.
For example, an extreme hexagonal morphology, $[6.6.6]$, is created by joining six SSPs together, as shown in \cref{fig:sspimplementation}(a).
The periodicity is indicated in the figure by matching colored lines along the outer edges.
Any two-dimensional periodic morphology can be constructed from SSPs in this way, though, as noted above, some morphologies require more than six triangles in an~\hbox{SSP}.

Given the locations of all the nodes, the stretching and interfacial energies of each SSP can be computed as described above.
The total free energy per chain for the periodic patch is then the sum of free energies of all the SSPs weighted by their areas, and divided by the total area of the patch.

The free energy does not depend on the location of the patch as a whole, so we choose one node, the reference node, to be fixed in space and make it the origin of the coordinate system.
This is indicated as a pink dot in \cref{fig:sspimplementation}(b).
The other nodes are either `free nodes' or `periodic nodes', indicated in green and light blue respectively in \cref{fig:sspimplementation}(b).
Free nodes can move anywhere in the 2D space, subject to the constraint that all SSPs remain valid (see Supporting Information).
Periodic nodes are those nodes on the edge of the patch whose location is constrained by the overall periodicity of the morphology and the locations of the free nodes.
The periodicity is specified by choosing two vectors (in two dimensions), $(v_{xx},0)$ and $(v_{xy},v_{yy})$ as indicated by dotted arrows in \cref{fig:sspimplementation}(b).
The first vector has a zero $y$~component in order to remove rotations, which don't affect the total free energy.
The locations of the periodic nodes are derived from the locations of the free nodes using integer sums of these two vectors.
For example, the position of a periodic node $(p_x, p_y)$  is related that of a free node $(x,y)$ by
\begin{equation}\label{ch4eq:periodicity equation}
	\begin{split}
		p_x&=x + l v_{xx} + m v_{xy},\\
		p_y&=y \phantom{{}+ l v_{xx}} {} + m v_{yy},
	\end{split}
\end{equation}
where $l$ and $m$ are integers.
Every periodic node is fixed in this way, relative to a particular free node.
\revision{The choice of assigning nodes as ``reference'', ``periodic'' or ``free'' is not unique, but the final minimized structure and free energy does not depend on this choice, apart from minor numerical differences that will arise in the minimization procedure.}

To take the hexagonal morphology shown in \cref{fig:sspimplementation}(b) as a specific example, the six vertices labelled $0$, $14$, $6$, $15$, $1$ and $8$ form one \hbox{SSP}, as indicated  by grey dotted lines.
(The numbering is different from that in \cref{ch1fig:sspframework}.)
Node~$0$ is the reference node.
The linked pairs of free nodes and periodic nodes are $[7,10]$, $[0,2]$, $[0,4]$,  $[8,11]$, $[1,3]$, $[1,5]$ and $[9,12]$.
This choice is not unique.
The seven nodes in the centre are not linked to any other nodes.
The configuration of the six SSPs is then specified by the two-dimensional locations of eleven free nodes and the three numbers $v_{xx}$, $v_{xy}$ and~$v_{yy}$ that specify the periodicity, making 25 degrees of freedom in all.
The free energy of the patch depends on these degrees of freedom and the polymer parameters.
Other morphologies are constructed in a similar way.

The total free energy per chain of a patch with a given morphology is minimized in two stages.
First, we affinely scale all node positions to find the approximate optimum periodicity lengths for the structure as a whole.
Second, we allow all node locations to vary in a manner consistent with the periodicity constraints (whilst still allowing further small variations in the periodicity).
The free energy, viewed as a function of all the degrees of freedom, is minimized using the Broyden--Fletcher--Goldfarb--Shanno (BFGS) algorithm~\cite{nocedal2006numerical}, as implemented in SciPy.

Using this framework, we compute phase diagrams for the ABC star structure, working out which morphology has the lowest free energy as a function of the composition parameters $(\phi_A,\phi_B,\phi_C)$, for different monomer interaction strengths.
While exploring the $(\phi_A,\phi_B,\phi_C)$ phase diagram, for each morphology we usually perform the first minimization with $\phi_A=\phi_B=\phi_C$, starting from the morphology as constructed by hand.
For other values of $(\phi_A, \phi_B, \phi_C)$, it is necessary to have a good initial guess for the configuration in order for the minimization to succeed.
One cause of the difficulty is that the location of the core is determined by the values of $(\phi_A, \phi_B, \phi_C)$ and the six polygon node locations, and it is possible that the minimization algorithm asks for the free energy for parameter values where the core is outside the SSP, which is invalid.
In order to provide good initial guesses, we `grow' the phase diagram from any initial point in $(\phi_A, \phi_B, \phi_C)$, working outwards in concentric rings and using neighboring converged configurations as initial guesses, repeating any failed minimizations with alternative nearby initial guesses.
This approach is better than the alternative of traversing the phase diagram in multiple linear cuts.

\section{Results}

\begin{figure}
\begin{center}
\includegraphics[width=0.9\linewidth]{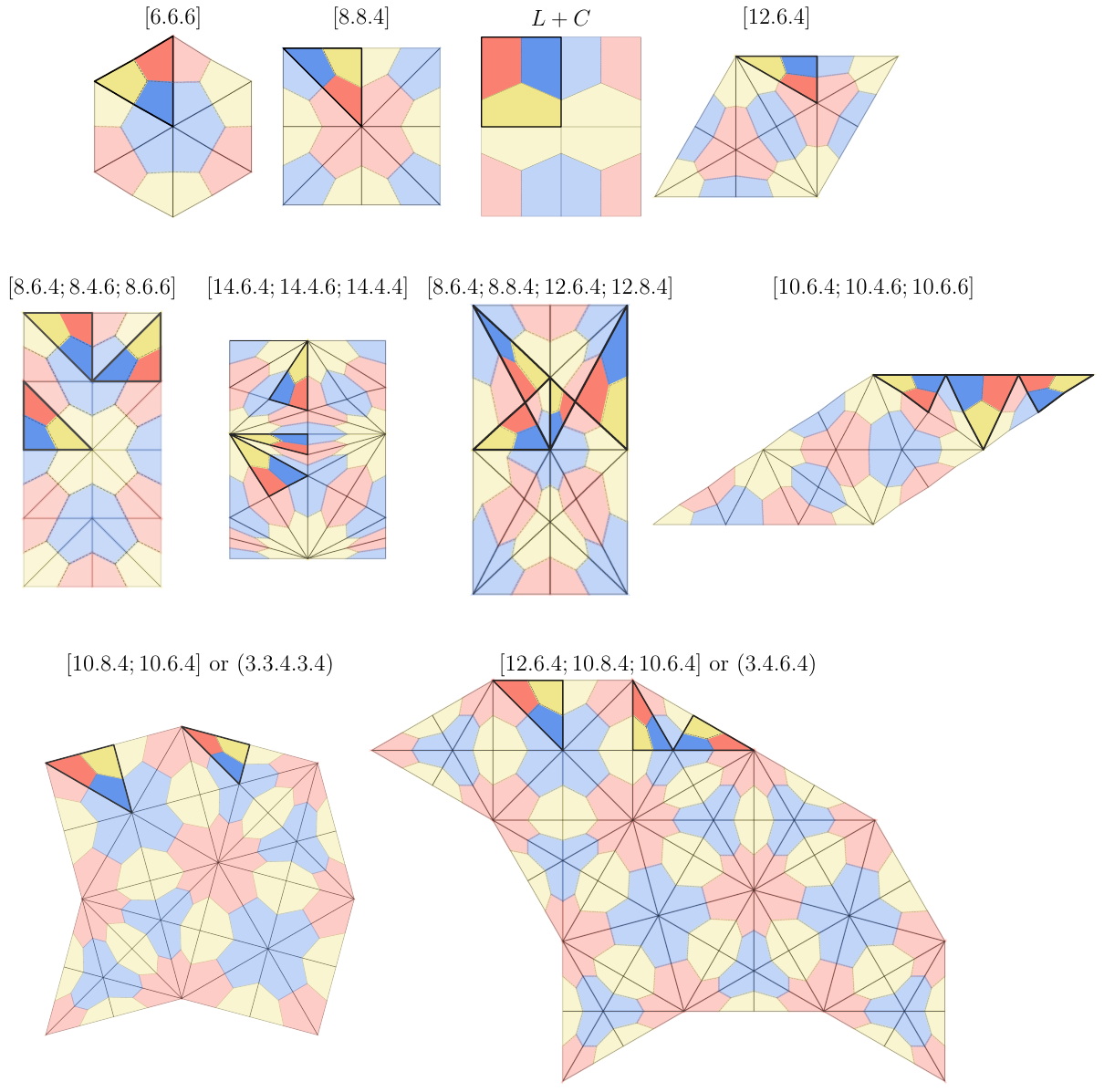}
\end{center}
\caption[Candidate morphologies]{\revision{This illustrates how the candidate morphologies considered in this work are discretized into SSPs. } 
The color scheme is A (red), B (blue) and C (yellow). 
The bold triangles are the SSPs, and each SSP contains six triangles and a single ABC core.
In these initial configurations, the SSPs are triangles but there are nodes at the domain boundaries on each edge, so they are in fact six-sided polygons (as in \cref{fig:sspimplementation}). 
The exception is the $L+C$ example, which has an eight-sided SSP containing eight triangles: two in the red and blue regions and four in the yellow.
The SSP polygons are different sizes, but the proportion of red, blue and yellow in each SSP is equal to $\phi_A$, $\phi_B$ and $\phi_C$, which in these examples are all equal to one third.  \revision{For each morphology we highlight, with darker shading, a single example of an SSP corresponding to each of the Schläfli symbols. So, for single core-type structures such as $[8.8.4]$ we highlight only one SSP. For multi-core structures several SSPs are highlighted, e.g., for $[8.6.4; 8.4.6;8.6.6]$, three SSPs are highlighted corresponding to examples of $[8.6.4]$, $[8.4.6]$ and $[8.6.6]$ cores.}
These initial configurations are the starting points for minimizing the free energy.}
  \label{fig:candidate_morphologies}
		\end{figure}

We will now use the SSP framework to demonstrate the analysis of different morphologies in ABC star terpolymers.
We explore both the structure (how the shapes of the domains change) and the stability (which morphology has the lowest free energy) of the phase-separated morphologies.
We will examine those single-core and multi-core morphologies that have been observed in experiments\cite{Okamoto1997MorphologyCopolymers,Huckstadt2000SynthesisPoly2-vinylpyridine,Takano2005,Chernyy2018Bulk} and predicted by theoretical work\cite{Gemma2002,Cody2024StableInteractions}.
The ten morphologies we consider in this paper are given in \cref{fig:candidate_morphologies} (see also \cref{fig:morphologies_withcore}).
In each case, we create the candidate morphologies by hand from existing patterns using drawing software, identifying the different monomer domains, selecting periodic vertices and calculating vertex locations from the drawing.
The spatially periodic region is divided into triangles (SSPs), such that each SSP contains exactly one ABC core. 
The locations of the vertices of the SSPs and the arrangement of the monomer types define the initial configuration of the SSPs for each morphology.

\subsection{Candidate morphologies}

In single-core morphologies, at the minimum free energy of the structure, all SSPs are rotations or reflections of each other throughout the periodic patch.
The single-core morphologies that have been observed in experiments are $[6.6.6]$, $[8.8.4]$, $[12.6.4]$ and the lamellar + cylindrical (L+C) morphology (see \cref{table:experimentmorphologies}). 
For the $[6.6.6]$ morphology, the periodic patch that we choose is a hexagon, with one domain (for example, B~in \cref{fig:candidate_morphologies}) surrounded by six other alternating domains of the other two monomer types. 
For the $[8.8.4]$ morphology, the periodic patch is a square, with one domain (A~in \cref{fig:candidate_morphologies}) surrounded by eight alternating domains of the other two monomer types.
For the $[12.6.4]$ morphology, the periodic patch is a rhombus. This rhombus is divided into two triangles, each one of them containing one monomer type in the middle (C~in \cref{fig:candidate_morphologies}), surrounded by six alternating domains of the other two monomer types.
The final single core morphology we consider is the L+C morphology, with one monomer type forming lamellae (B~in \cref{fig:candidate_morphologies}) with the other two monomer types alternating on either side.
Unlike the first three morphologies described here, L+C requires a geometrically asymmetric SSP with eight sides, due to the connectivity of lamellae within this morphology. 

In multi-core morphologies, the \revision{SSPs} need not all be the same throughout the patch.
The multi-core morphologies that we consider are
$[8.6.4; 8.4.6; 8.6.6]$,
$[14.6.4; 14.4.6; 14.4.4]$,
$[8.6.4; 8.8.4; 12.6.4; 12.8.4]$, 
$[10.6.4; 10.4.6; 10.6.6]$, 
$[10.8.4; 10.6.4]$ and
$[12.6.4; 10.8.4; 10.6.4]$.
These are all considerably more complicated to construct.
In practice, after identifying the periodic patch in each case, we identified the locations of all the cores and divided the periodic patch into triangles.
We overlaid SSPs onto these triangles, and took the resulting coordinates as our initial configuration.  \revision{For each structure in \cref{fig:candidate_morphologies} we highlight one example of an SSP that corresponds to each of the number triplets in the Schläfli symbols, e.g., for $[8.6.4; 8.4.6;8.6.6]$, three SSPs are highlighted, corresponding to examples of $[8.6.4]$, $[8.4.6]$ and $[8.6.6]$ cores.  Here it is important to note that not all cores with the same number triplets are equivalent by symmetry, and the local environment of their neighbouring SSPs may be different.  One consequence, which we shall detail below, is that even if two SSPs appear identical in the starting structures of \cref{fig:candidate_morphologies}, their shape after free energy minimization can be different.}

In addition, each morphology comes in six variants, depending on how the monomer types are allocated to the SSP triangles.

\subsection{Construction of phase diagrams}

We choose values of $N\chi_{AB}$, $N\chi_{BC}$ and $N\chi_{AC}$ and construct ternary phase diagrams with the above-mentioned morphologies with monomer compositions $(\phi_A, \phi_B, \phi_C)$ on each axis.
We vary monomer compositions from $0$ to $1$ with an increment of $0.01$, with $\phi_A+\phi_B+\phi_C=1$.
The free energy minimization struggles when one polymer type dominates, so we restrict to $0.05\leq\phi_A\leq0.90$, $0.05\leq\phi_B\leq0.90$, $0.05\leq\phi_C\leq0.90$, resulting in around 4000 distinct points in the ternary phase diagram.
The value of the minimized free energy per chain for each morphology is compared at every point in this diagram, and the morphology with the lowest free energy per chain is marked at that point.  

For any given tiling pattern it is possible to produce six different structures by permutation of the monomer ``colors'' (i.e., red, yellow, blue in our figures).
For a given monomer composition and set of interaction parameters, each permutation will have a different free energy.
The free energy maps for each of the candidate morphologies are created for all permutations of the domains (ABC, ACB, BAC, BCA, CAB, CBA). 
The final phase diagram is created by overlaying the free energy maps for all six permutations for the ten candidate morphologies.
The morphology with lowest free energy is the one that is marked in the final diagram.

\revision{As noted above, the degree of polymerization~$N$ affects the parameter~$c$ in the expression for the stretching free energy. This changes the balance between stretching and interfacial energies, and so affects the phase diagram.
For most phase diagrams in this paper, we set the degree of polymerization $N=1000$.
We also include phase diagrams with $N=300$ and $N=10000$ in the Supporting Information and discuss the effect of varying~$N$ below.}

\begin{figure}
\begin{center}
\includegraphics[width=1\linewidth]{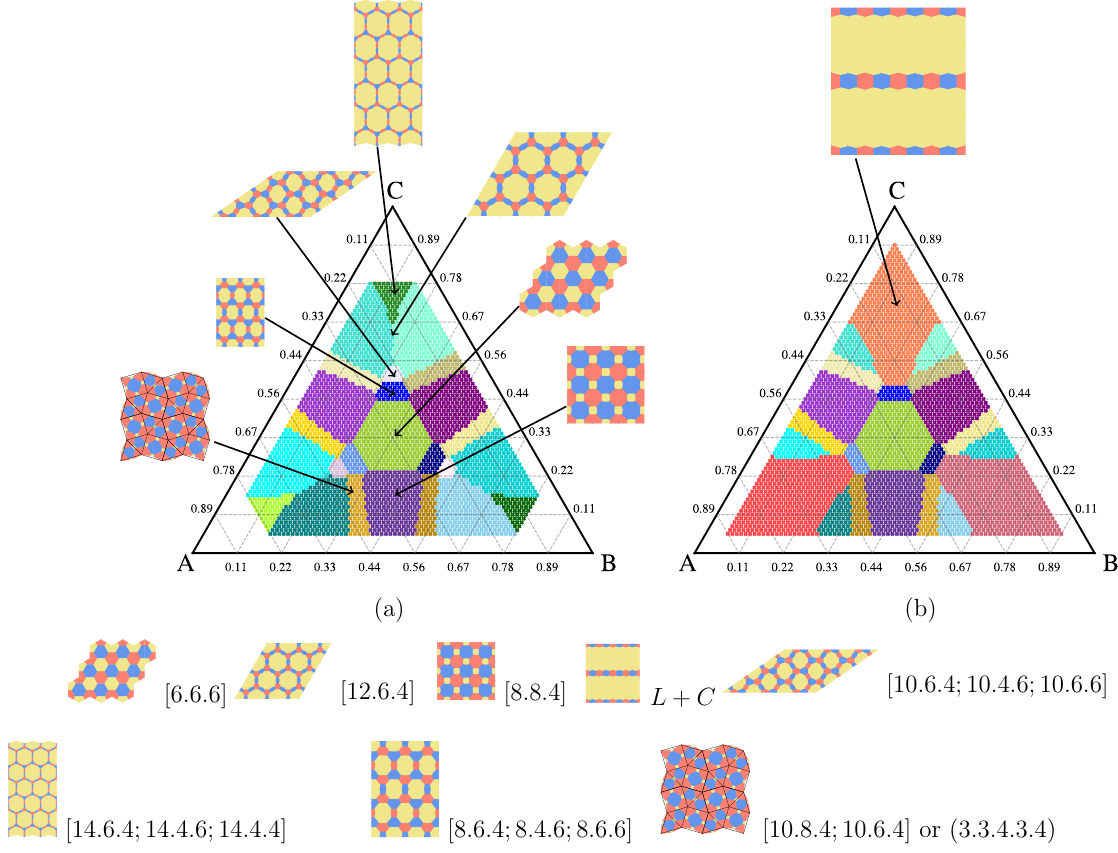}
		\end{center}
		  \caption[Phase diagram symmetric]{Phase diagrams for ABC star terpolymer with symmetric interaction strengths ($N\chi_{AB}=N\chi_{BC}=N\chi_{AC}=60$) and with \revision{$N=1000$}.
          On the left~(a) is the phase diagram created using only six-sided SSPs. 
          The stable morphologies are marked around the phase diagram, with the six permutations of each morphology indicated with shades of color.
          On the right~(b) is the phase diagram where we include eight-sided SSPs, which allow lamellar phases. The data for this figure is available from~\cite{Joseph2025data}.}
  \label{fig:symmetric_phase_diagram}
		\end{figure}

\subsection{ABC star terpolymers with symmetric interactions}

The phase diagrams for ABC star terpolymer melts with symmetric interaction strengths ($N\chi_{AB}=N\chi_{BC}=N\chi_{AC}=60$) and with \revision{$N=1000$} is given in \cref{fig:symmetric_phase_diagram}.
The value of~$N\chi=60$ is chosen in order to be in the strong segregation limit~\cite{LiW2010}.
This gives $\gamma_{AB}=\gamma_{BC}=\gamma_{AC}=\sqrt{10}$ in \cref{eq:interfacial_energy_per_chain_polygon}.
We vary from this symmetric case in the next section.
As expected, the phase diagrams have mirror symmetry in the lines $\phi_A=\phi_B$, $\phi_B=\phi_C$ and $\phi_A=\phi_C$ in the ternary space when we do not distinguish between the color permutations within each morphology.

All of our chosen candidate morphologies in \cref{fig:candidate_morphologies} are found in these phase diagrams apart from $[8.6.4;8.8.4;12.6.4;12.8.4]$  and $[12.6.4; 10.8.4; 10.6.4]$. 
All morphologies that are experimentally observed in ABC terpolymer systems~\cref{table:experimentmorphologies} are present in the phase diagram.

In \cref{fig:symmetric_phase_diagram}(a), we show all morphologies apart from the lamellar phases, and in \cref{fig:symmetric_phase_diagram}(b), we include the lamellar phases as well.
These lamellar phases, indicated with orange shades, appear in the corners of the phase diagram.
They overtake the entire regions of $[10.6.4;10.4.6;10.6.6]$ and $[14.6.4;14.4.6;14.4.4]$ phases, and the regions of $[12.6.4]$ are much reduced.
The multi-core structures that are not obscured are $[8.6.4;8.4.6;8.6.6]$ and the $\Sigma$-phase. 
The reason we include both versions of the phase diagram is that in phase diagrams computed using self-consistent field theory~\cite{Cody2024StableInteractions}, the lamellar phases are only found in the very corners of the phase diagram.
In addition, the Monte Carlo calculations of Gemma {\textit et al.}\cite{Gemma2002} have $[12.6.4]$ phases along the axis of symmetry connecting the corner to the center of the diagram, and again this region is obscured by the lamellar phase.
As a result, it is interesting to know what lies underneath the lamellar phases in the central parts of the phase diagram in our strongly-segregated calculations.
All parts of these diagrams are computed using six-sided SSPs, apart from the lamellar phases in~(b), which require eight-sided~SSPs. 

We first discuss the regions with single-core structures in the phase diagrams in \cref{fig:symmetric_phase_diagram}.
The light green region in the center is where the $[6.6.6]$ morphology has the lowest free energy per chain.
This occurs when all three branches have comparable lengths, $\phi_A\approx\phi_B\approx\phi_C$.
The hexagonal green region where the $[6.6.6]$ morphology has the lowest free energy is bordered by three regions of $[8.8.4]$ (purple shades) and three regions of $[8.6.4;8.4.6;8.6.6]$ (blue shades).
The $[8.8.4]$ morphology has the lowest free energy when two branches have comparable lengths and the third is smaller than the other two (for example, $\phi_A\approx\phi_B>\phi_C$).
The other single-core structures ($[12.6.4]$) are indicated in teal/light blue shades, and they occupy the corner regions of the ternary space, where all three chains have different lengths, with one chain significantly longer than the other two (for example, $\phi_A>0.5>\phi_B>\phi_C$).
It is interesting to observe that single-core structures occupy the majority of the phase diagram.

Next we turn to regions with multi-core structures, which are stable at the intersections of the $[6.6.6]$, $[8.8.4]$ and $[12.6.4]$ single-core regions.
Between the $[6.6.6]$ region in the center and the six outer $[12.6.4]$ regions, we find (moving outwards) $[8.6.4;8.4.6;8.6.6]$ (trapezoidal shaped regions indicated with blue shades) and $[10.6.4;10.4.6;10.6.6]$ (triangular shaped regions 
indicated with lavender shades).
In all these regions, the compositions of two of the polymer types are comparable, while the third is somewhat longer than the other two: none of the three branches are extremely small.
Towards the corners of the phase diagram, and at the intersection of regions occupied by two topological permutations of $[12.6.4]$, there are regions of $[14.6.4;14.4.6;14.4.4]$ (indicated with shades of dark green).
Like the $[12.6.4]$ structures, this morphology has a large number of cores surrounding one central domain, and in these regions, the compositions of two of the polymer types are small and comparable, while the third is considerably longer than the other two.

The third type of multi-core structure that we find is the $[10.8.4;10.6.4]$, also known as the $\Sigma$-phase and $(3.3.4.3.4)$.
The six rectangular-shaped regions of this phase (indicated with yellow shades) are found between the (purple) $[8.8.4]$ and (teal) $[12.6.4]$ regions.
The $[8.8.4]$ phase comprises square tiles (made from joining the centers of the blue, red or yellow domains), and the $[12.6.4]$ phase comprises triangular tiles (made from joining the centers of the largest of the three colors).
Since the $\Sigma$-phase is a combination of square tiles and triangular tiles, it is natural to find this phase between the regions of $[8.8.4]$ and $[12.6.4]$. 
In the phase diagram given in \cref{fig:symmetric_phase_diagram}(a),  all the rectangular regions corresponding to different topological sub-classes of the $\Sigma$-phase occupy equal areas. 
In these calculations using symmetric interactions, we find the $\Sigma$-phase at roughly the same position in the phase diagram (using our SSP method) as reported by Ueda et al.~\citep{Ueda2007} (using Monte Carlo methods) and by Li et al.~\citep{LiW2010}, Xu et al.~\cite{Xu2013ACopolymers} and Cody et al.~\cite{Cody2024StableInteractions} (using SCFT). 
While the rough locations of the regions of $\Sigma$-phase are similar, our regions are somewhat larger than those reported in the literature.

In the Supporting Information, we show phase diagrams for \revision{$N=300$ and $N=10,000$}.
These are qualitatively similar to the \revision{$N=1000$} phase diagrams presented here.
\revision{The most important effect of changing~$N$ is to change the extent of the L+C phases.
With smaller $N=300$, the regions occupied by the L+C phases are larger, and as a consequence, the regions occupied by $[12.6.4]$ are reduced.
With larger $N=10,000$, the regions occupied by the L+C phases are smaller, revealing the previously hidden $[10.6.4;10.4.6;10.6.6]$ structures.}
There are also small changes in the positions of the boundaries between other phases, but without a qualitative change to the overall phase diagram.

\revision{
The phase diagrams shown in \cref{fig:symmetric_phase_diagram} agree with ternary phase diagrams for ABC terpolymer systems reported by Hawthorne et al.,~ \cite{hawthorne_stable_2024} Li et al.~\cite{LiW2010}, and Zhang et al.~\cite{Zhang2010}, which were obtained using SCFT for 2D morphologies.
Overall, the types of morphologies observed are qualitatively similar; however, a notable difference arises for the L+C phase.
In prior studies, the L+C and other lamellar phases occupy relatively limited regions of the phase space, whereas in our phase diagram (\cref{fig:symmetric_phase_diagram}.(b)) L+C occupies a substantial portion of the diagram.
Multicore morphologies, $[10.6.4;10.4.6;10.6.6]$ and $(3.4.6.4)$, which are reported in earlier work, are absent in our results. As $N$ increases, some multicore morphologies reappear, and the resulting phase diagrams progressively resembles those predicted by SCFT. 
}

\subsection{ABC star terpolymers with asymmetric interactions}

The \hbox{SSP} technique allows us to vary parameters with relative ease and so we can readily study morphologies with asymmetric interaction strengths.
We start from the symmetric phase diagrams in \cref{fig:symmetric_phase_diagram}, with $(N\chi_{AB},N\chi_{BC},N\chi_{AC})=(60,60,60)$ with \revision{$N=1000$}, and recalculate all minimized free energies with different values of $(N\chi_{AB},N\chi_{BC},N\chi_{AC})$.

\begin{figure}
\begin{center}
\includegraphics[width=1\linewidth]{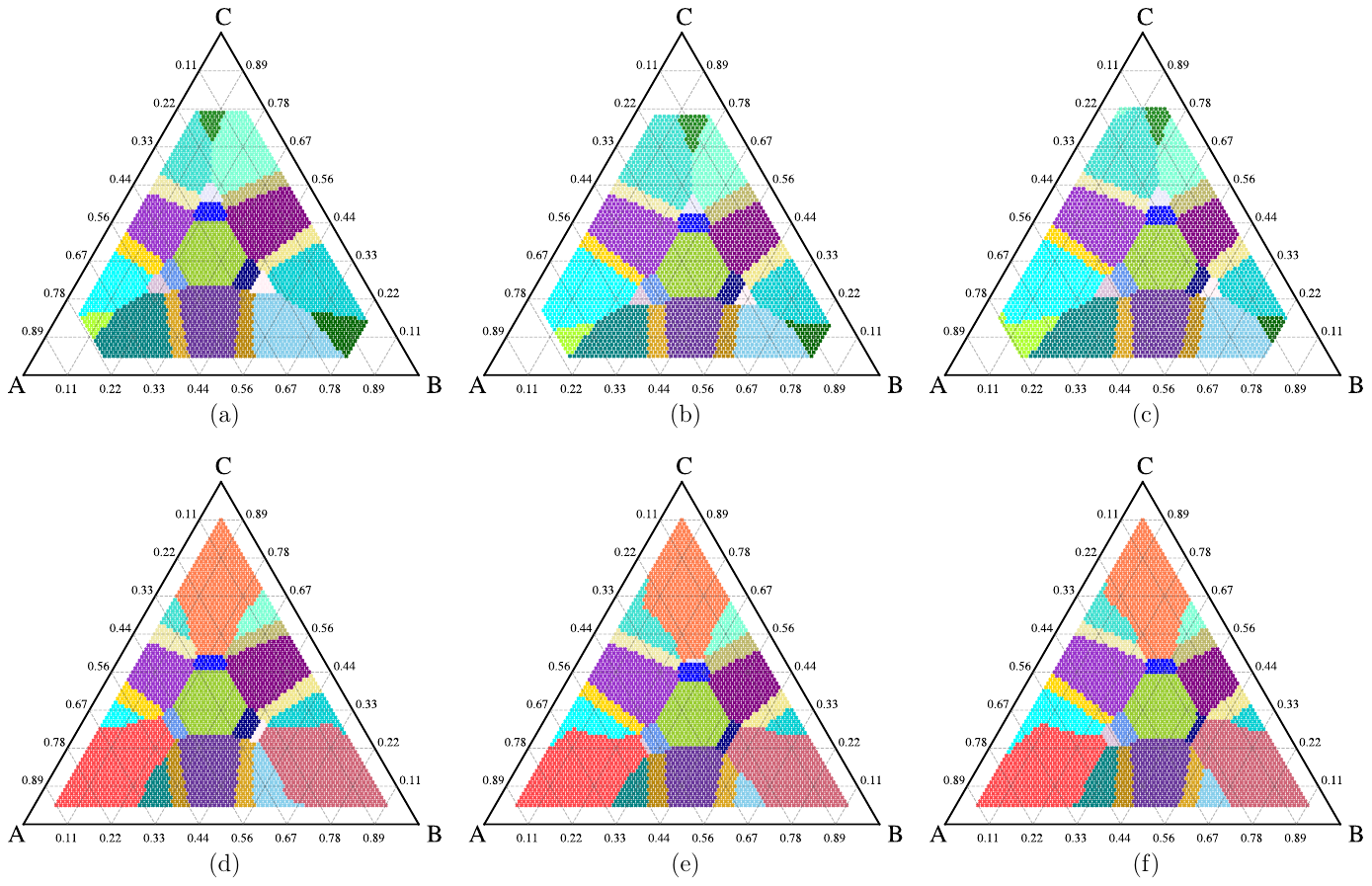}
		\end{center}
\caption[Asymmetric phase diagrams]{Phase diagrams of ABC star terpolymers with asymmetric interactions and with \revision{$N=1000$}. 
In (a) and~(d), \revision{$N\chi_{AB}=N\chi_{BC}=120$, $N\chi_{AC}=60$}.
In (b) and~(e), \revision{ $N\chi_{AB}=N\chi_{BC}=60$, $N\chi_{AC}=120$}.
In (c) and~(f), \revision{$N\chi_{AB}=120$, $N\chi_{BC}=60$, $N\chi_{AC}=180$}.
The top row has phase diagrams with morphologies using only six-sided SSPs.
The bottom row has phase diagrams that include lamellar morphologies, constructed using eight-sided SSPs. 
The colors indicating the morphologies are same as in the symmetric phase diagrams in \cref{fig:symmetric_phase_diagram}.
The data for this figure is available from~\cite{Joseph2025data}.}
  \label{fig:asymmetric_phase_diagram}
\end{figure}

In \cref{fig:asymmetric_phase_diagram}, we report three cases of different asymmetric interactions. The three cases are:
\begin{enumerate}

\item \revision{$N\chi_{AB}=N\chi_{BC}=120$, $N\chi_{AC}=60$, so there is less repulsion between A and C monomers compared to the repulsion between the other two pairs, which are equal.}

\item \revision{ $N\chi_{AB}=N\chi_{BC}=60$, $N\chi_{AC}=120$, so there is more repulsion between A and C monomers compared to the repulsion between the other two pairs, which are equal.}

\item \revision{$N\chi_{AB}=120$, $N\chi_{BC}=60$, $N\chi_{AC}=180$, so B and C repel each other less strongly, and A and C repel each other more strongly, compared to the interaction between A and~\hbox{B}.}
\end{enumerate}
Phase diagrams for these three cases using the six-sided SSPs with \revision{$N=1000$} are given in \cref{fig:asymmetric_phase_diagram}(a), (b) and~(c), and the phase diagrams including eight-sided SSPs are given in  \cref{fig:asymmetric_phase_diagram}(d), (e) and~(f).

As expected, we observe shifts in the phase diagrams relative to the symmetric case.
For the cases 1 and 2, given in \cref{fig:asymmetric_phase_diagram}(a,d) and \cref{fig:asymmetric_phase_diagram}(b,e), there is a mirror symmetry with respect to the diagonal line $\phi_A=\phi_C$, as expected from having $N\chi_{AB}=N\chi_{BC}$.
This symmetry is not present when all interactions are different. 
The overall phase diagram is similar to the symmetric case, with same relative positions of the different regions in the phase diagram, but the sizes of the regions change and there is an overall shift. 
In the case with $N\chi_{AB}=N\chi_{BC}$, this shift is along the line of symmetry ($\phi_A=\phi_C$), with all regions moving away from the B~corner when $N\chi_{AC}<N\chi_{AB}$ and moving towards the B~corner when $N\chi_{AC}>N\chi_{AB}$.

With changing the values of the $N\chi$ parameters, we observe that regions occupied by different topological sub-classes within the same morphology are no longer identical.
\revision{For example, in \cref{fig:asymmetric_phase_diagram}(a,d), the yellow regions corresponding to the stable $\Sigma$-phase and located farther from the B corner are wider than the other four regions. In contrast, in \cref{fig:asymmetric_phase_diagram}(b,e) and (c,f), the yellow regions farther from the B corner are thinner than the other four stable $\Sigma$-phase regions in their respective phase diagrams.}

The $[10.6.4;10.4.6;10.6.6]$ regions (lavender shades) were masked by the lamellar phases (orange shades) in the symmetric case, but with asymmetric interactions, decreasing $N\chi_{AC}$ reveals those regions with larger B domains, and vice versa.
In the more extreme example in \cref{fig:asymmetric_phase_diagram}(f), the $[10.6.4;10.4.6;10.6.6]$ region closest to the A corner is \revision{partially} revealed with the small value of $N\chi_{BC}$, while the $[8.6.4;8.4.6;8.6.6]$ region (blue shades) closest the the B~corner (as well as the $[10.6.4;10.4.6;10.6.6]$ region) is completely masked by the lamellar phase.

\begin{figure}
    \centering
    \includegraphics[width=1.0\linewidth]{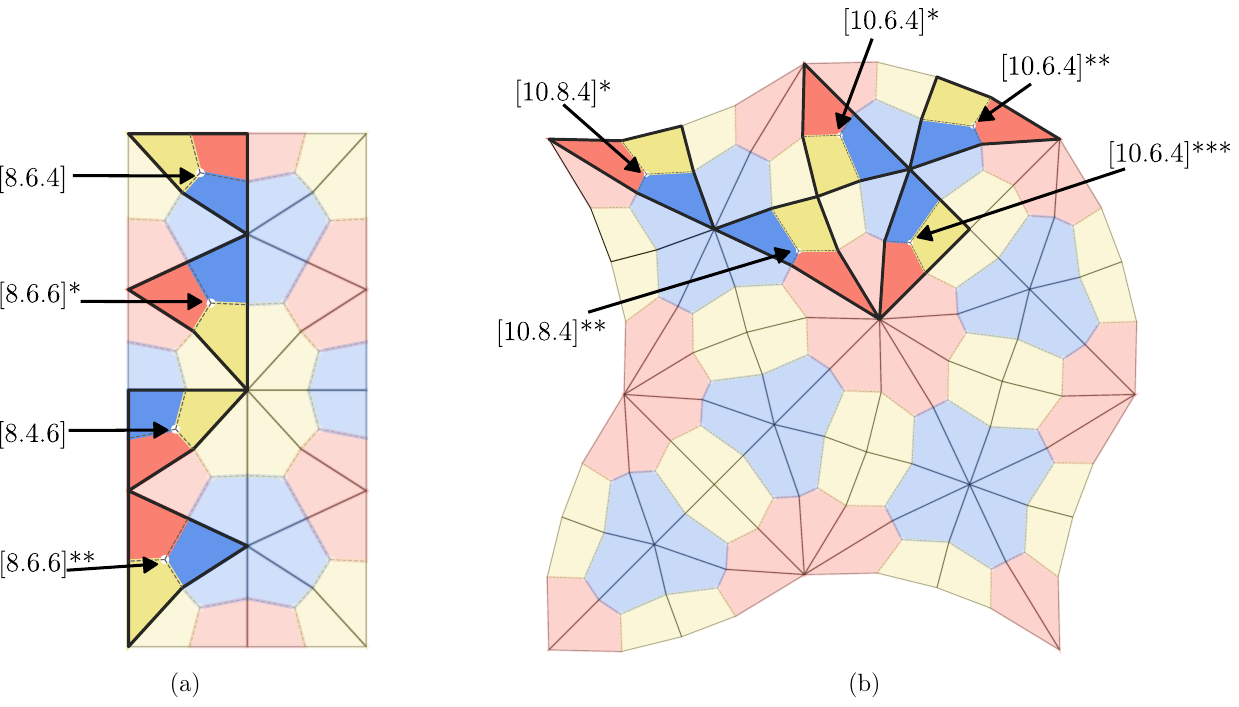}
    \caption{\revision{Periodic patches of morphologies $[8.6.4;8.4.6;8.6.6]$ and $\Sigma$-phase after minimisation are shown with highlighted SSPs. Examples of each symmetrically distinguishable SSPs are highlighted. In~(a), $[8.4.6]$ and $[8.6.4]$ are highlighted as well as two types of $[8.6.6]$.
    In~(b), two types of $[10.8.4]$ and three types of $[10.6.4]$ are highlighted. } }
    \label{fig:ssps_after_minimisation}
\end{figure}

\subsection{Structural analysis of minimized morphologies}

In the existing literature on polymer phase separation, there is little discussion of the shape of local monomer domains and the curvature of their interfaces.
Our SSP method allows us to visualize the domains and investigate the interfaces between them, both at the scale of the local monomer domains and at the scale of the tiles that make up the morphology.

\revision{We first examine the shapes of SSPs in the multicore structures.
As noted above, within each structure, SSPs with the same Schläfli symbol number triplets are not necessarily equivalent by symmetry, having different local environments.
As illustrated in \cref{fig:ssps_after_minimisation}, this affects their area and shape in the structures after free energy minimization.
For the $[8.6.4;8.4.6;8.6.6]$ morphology shown in \cref{fig:ssps_after_minimisation}(a), all $[8.6.4]$ SSPs are symmetrically equivalent and so take the same shape upon minimization; likewise for the $[8.4.6]$ SSPs.
However, the $[8.6.6]$ SSPs fall into two classes, marked as $[8.6.6]^{*}$ (with a concave red--yellow edge) and $[8.6.6]^{**}$ (with a concave blue--yellow edge) in the figure. 
Similarly, for the $\Sigma$-phase in \cref{fig:ssps_after_minimisation}(b), there are two classes of $[10.8.4]$ and three classes of $[10.6.4]$ SSPs, each with different shapes in the minimised structure.}

\begin{figure}
\begin{center}
\includegraphics[width=1\linewidth]{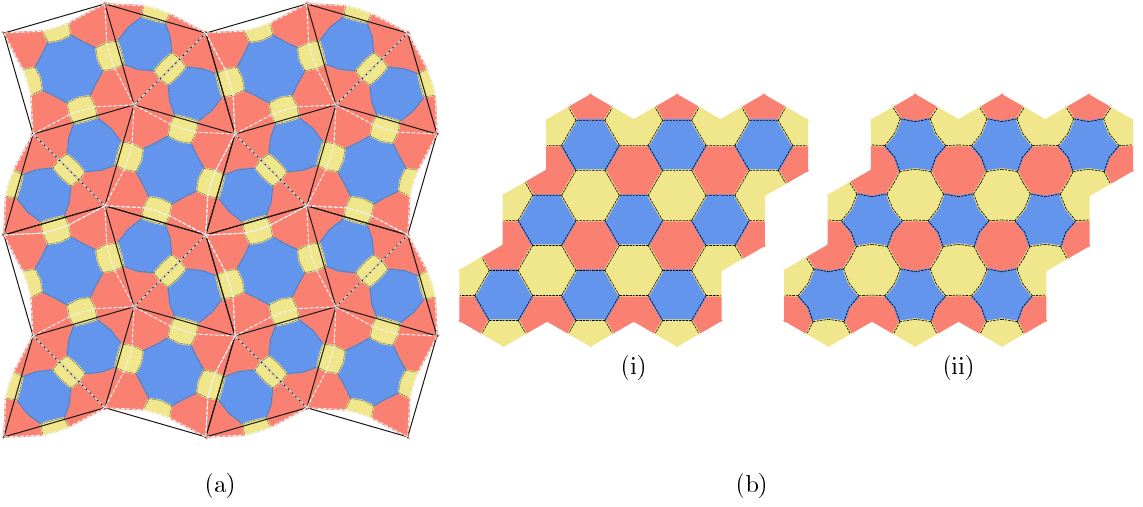}
\end{center}
\caption[Minimized structures]{Minimized structure for (a)~the $\Sigma$-phase and (b)~$[6.6.6]$ with different interaction strengths.
In~(a), we have $\phi_A=0.57$ (red), $\phi_B=0.38$ (blue), $\phi_C=0.05$ (yellow) and $N\chi_{AB}=N\chi_{BC}=N\chi_{AC}=60$.
The original (unminimized) tile edges are straight black lines, and the tile edges after minimization are shown as dashed white lines
In~(b), we have $\phi_A=\phi_B=\phi_C=\frac{1}{3}$, and in~(i), $N\chi_{AB}=N\chi_{BC}=N\chi_{AC}=60$, 
while in~(ii),\revision{ $N\chi_{AB}=120$, $N\chi_{BC}=60$ and $N\chi_{AC}=180$}.}
\label{fig:structure_morphology}
\end{figure}

\revision{As we noted in the introduction, it is helpful for construction of the morphology to view the $\Sigma$-phase as being composed of square and triangular tiles. At the level of these} tile shapes, the initial configuration of the $\Sigma$-phase in \cref{fig:candidate_morphologies} contains two square and four equilateral triangle tiles, all with with straight edges.
Upon minimization, we obtain the morphology state given in \cref{fig:structure_morphology}(a).
There are significant changes in the tile shapes: all edges of the square tiles curve inwards, and two edges of every triangle tile curve outwards.
In \cref{fig:structure_morphology}(a), the original (unminimized) tile edges are straight black lines, and the tile edges after minimization are shown as dashed white lines.
The edges between two adjacent triangles are straight, while the edges between triangle and square tiles curve towards the square tile.
This is seen most clearly in the locations of the yellow domains, which are centred on the edges between two triangles, but are shifted towards the squares when they are between a triangle and a square.
We observe the same effect for all value of the monomer compositions and interaction strengths, and for all morphologies that involve both square and triangle tiles.

Zeng et al.~\cite{Zeng2023ACells} attribute the changes of the tile shapes (in a related system) to the difference in the number of core per unit area between the tiles:
with edges of unit length, squares (with 8~cores) have 8 cores per unit area, while triangles (with 6~cores) have $24/\sqrt{3}\approx13.9$ cores per unit area.
Balancing the stretching energies between these two requires squares to shrink and triangles to expand.
This effect is seen in SCFT calculations \cite{LiW2010,Cody2024StableInteractions} and in the experimental results reported by Hayashida et al.~\cite{Hayashida2007a}, but was not remarked.

At the level of the local monomer domains, our SSP method provides information about the shapes of the interfaces between these domains.
In \cref{fig:structure_morphology}(b), we show how changing the interaction strengths affects the shapes of the domains.
In~(i), with $\phi_A=\phi_B=\phi_C=\frac{1}{3}$ and $N\chi_{AB}=N\chi_{BC}=N\chi_{AC}=60$, the monomers are all equivalent, and the morphology with the lowest free energy is perfect hexagons, with all monomer domains having the same area.
In~(ii), we change the interaction strengths and set \revision{$N\chi_{AB}=120$, $N\chi_{BC}=60$ and $N\chi_{AC}=180$}, resulting in curved interfaces while keeping all the areas the same.
Since $N\chi_{AC}$ is the largest, the incompatibility between A and C domains is minimized by reducing the length of the AC (red--yellow) interfaces.
Similarly, $N\chi_{BC}$ is the smallest, resulting in longer BC (blue--yellow) interfaces.
The corners of the A domains are all attached to BC interfaces, which can move most easily, resulting more circular A domains, with AB and AC interfaces curving outwards from the A domain.
Conversely, the corners of B domains are pulled by the shorter AC interfaces, resulting in pointed corners in B domains, with AB and BC interfaces curving into the B~domain.

We illustrated this effect with the equal-area $[6.6.6]$ morphology, but similar arguments apply to other morphologies.  \revision{Higher interactions (higher surface tensions) tend to favour more circular domains.  This general principle helps to rationalise the shifts in the phase diagram observed above in \cref{fig:asymmetric_phase_diagram}. Domains that have higher numbers of neighbours are able more easily to approximate a circle. Hence, in regions of the phase diagram where two morphologies are competing, higher interaction strengths for a given monomer type will tend to favour morphologies where the number of neighbours is larger for domains of those monomers.  This can be observed, for example, in \cref{fig:asymmetric_phase_diagram}(a), where $N\chi_{AB}=N\chi_{BC}=120$, $N\chi_{AC}=60$, so that the interactions involving the B~monomers are largest.  Here one can observe that the $[12.6.4]$ phases with 12-neighbour B~domains displace the $[10.8.4; 10.6.4]$ $\Sigma$-phases with 10-neighbour B domains, which in turn displace the $[8.8.4]$ phases with 8-neighbour B~domains.  These in turn displace the $[6.6.6]$ phase, which (finally) displaces the $[8.8.4]$ phases with 4-neighbour B~domains.  In every case, the morphology where the B-domain has the larger number of neighbours ``wins''.  The net effect is that the overall phase diagram appears to shift away from the B~corner towards the A-C boundary.  An exception to this shifting trend is the L+C morphology. For example, when B-domains are lamellar they cannot approximate a circle, and so this morphology does not increasingly outcompete other phases when the B interactions are increased. Hence, while other phases shift in the phase diagram, the L+C phase does not, which is why we see the emergence of the $[10.6.4; 10.4.6; 10.6.6]$ phase towards the bottom right of \cref{fig:asymmetric_phase_diagram}(d) as it outcompetes the L+C phase on the one side and $[8.6.4; 8.4.6; 8.6.6]$ on the other.}

\section{Conclusions}

We have developed a method for investigating different morphologies in ABC star terpolymers in the strong segregation limit.
Our method is based on surrounding each ABC core with a polygon, and then adjusting the locations of the vertices of all the polygons to find an overall minimum of the free energy.
We mainly use six-sided polygons, and eight-sided polygons for certain morphologies, but the method can be extended to polygons with more sides if needed (for example, if greater resolution of interfacial shapes was required).
This improves on the method of Gemma~\cite{Gemma2002}, who did strong segregation calculations of ABC star terpolymers using triangles rather than polygons around each core.
Other strong segregation calculations of AB diblock~\cite{Grason2003, Olmsted1998} and ABC tri-block~\cite{Phan1998} primarily used wedges or other elements to parameterize the shapes of the regions.
For ABC star terpolymers, our polygons are the simplest way of describing the phase separated structures, and so our work is a significant addition to the strong segregation theory toolbox.
We have considered here only two-dimensional tiling-based periodic structures; in principle, the method can be extended to consider periodic approximants to aperiodic (quasicrystalline) structures.

Apart from the work of Gemma~\cite{Gemma2002}, there are no other strong segregation calculations for these kinds of ABC star structures in the literature, though there are many self-consistent field theory calculations of these structures and more.
Strong segregation theory calculations are much more straight-forward than self-consistent field theory and other mean field theory calculations.
Using strong segregation theory allows rapid exploration of many different structures in the phase diagram.
Using strong segregation theory, combined with our approach of `growing' phase diagrams from a converged structure, enables us to compute entire phase diagrams with much less computational effort than alternative methodologies (self-consistent field theory, dissipative particle dynamics and Monte Carlo).

In particular, we have computed phase diagrams for all the known two-dimensional structures in ABC star terpolymer systems, with both equal and unequal Flory interaction parameters, considering a wider range of asymmetric variations than elsewhere in the literature.
We observe that single core morphologies ($[6.6.6]$, $[8.8.4]$, $[12.6.4]$ and L+C) occupy the majority of all the phase diagrams, with the multi-core $\Sigma$-phase and $[8.6.4; 8.4.6; 8.6.6]$ also present.  While we focused on the \revision{$N=1000$ case, our $N=300$ and $N=10,000$ phase diagrams in the Supporting Information show that the regions of L+C are reduced and the multi-core $[10.6.4;10.4.6;10.6.6]$ morphology that was masked by L+C can be revealed with larger~$N$.}
This limit is appropriate for SCFT calculations, which find the same multi-core structure and a reduced prominence for lamellar phases.
Experiments with pure ABC star terpolymers record only single-core structures (see \cref{table:experimentmorphologies}, where structures featuring lamellae are prominent).
With unequal interaction parameters, we observe a shift in the phase diagram consistent with the self-consistent field theory calculations of Jiang et al.~\cite{Jiang2015} \revision{This shift in the phase diagram also allows the multi-core $[10.6.4;10.4.6;10.6.6]$ morphology to be revealed from behind the L+C morphology.}
More widely, similar shifts in phase diagrams are observed in diblock and other systems once asymmetry is introduced, which can reveal more complex phases such as Frank--Kasper phases~\cite{Grason2003}.

\revision{We were able to rationalize the shifts in the phase diagram with varying interaction parameters by noting that higher interactions gives higher interfacial tension and so a preference for more circular domains with a larger number of neighbours. The dependence of phase diagrams on $N$ (independently of the product $N \chi$) arises because of the logarithmic dependence of the stretching energy on the core radius $R_{core}$ in \cref{eq:Ball_stretching}.  Thus, changing $N$ affects the balance between stretching and interfacial energies, giving small shifts in the phase boundaries.}

The method can be extended in a variety of ways.  For instance, the effects of conformational asymmetry can be introduced by allowing the step length parameter~$b$ to be different for the different monomer types.
The method could be used to investigate mixtures of more than one type of terpolymer, for example, mixtures of ABC and ABD terpolymers, or mixtures of ABC terpolymer and A~homopolymer (as used experimentally by Hayashida et al.\cite{Hayashida2007a}). 

\revision{Other potential extensions to the method involve modifications to the SSP structure. In this paper we have used SSPs with a minimal number of nodes to represent the morphologies we have investigated, using six-sided SSPs for the majority of the morphologies and eight-sided SSPs for the L+C phase. Adding extra nodes would permit a more refined calculation of both structure and free energy, at the computational expense of an increased number of degrees of freedom.  Extra nodes along the edges of the SSP would permit a higher-order representation of the boundaries between SSPs where these are not straight, which occurs only in the multicore structures as illustrated in \cref{fig:ssps_after_minimisation}.  One could also consider adding nodes to obtain a higher-order approximation of the lines from the central node to the exterior of the SSP in \cref{ch1fig:sspframework}, allowing a better representation of curved interfaces between domains and, correspondingly, non-straight wedges along which the chains stretch.}

We have only implemented the method in two dimensions, though in principle it could be extended to consider three-dimensional structures. \revision{This would naturally involve extending each SSP into the third dimension, so that they comprise two hexagonal faces (each essentially identical to our 2D SSPs) connected by lines joining the equivalent outer and central nodes from the two faces.  For maximal flexibility, the triangles within the hexagonal faces need not be coplanar in 3D, and nor do the two hexagonal faces need to be parallel.  By stacking together a number of such base motifs, it will possible to approximate effects that arise in three dimensional structures, such as core lines that are not straight.}
However, almost all ABC star terpolymer structures that have been reported experimentally are two-dimensional (see \cref{table:experimentmorphologies}).

We also believe that it is possible to construct similar tessellating base motifs to perform SST calculations for other copolymer architectures.  \revision{As noted in the Introduction for example, Reddy, Dimitriyev and Grason~\cite{Grason2022, Grason2023} constructed a strong segregation theory for bicontinuous phases of AB copolymers by making use of ``wedges'',~\cite{Olmsted1998} which are the tessellating base motif structure for an AB copolymer, equivalent to our SSPs.}

In future, we plan to use this method to investigate quasicrystal approximants in ABC star terpolymers, inspired by the tilings presented by Wang et al.\cite{Cody2024StableInteractions}
The efficiency of the method will allow us to reach reasonably large approximants.

% \bibliography{references}

\providecommand{\latin}[1]{#1}
\makeatletter
\providecommand{\doi}
  {\begingroup\let\do\@makeother\dospecials
  \catcode`\{=1 \catcode`\}=2 \doi@aux}
\providecommand{\doi@aux}[1]{\endgroup\texttt{#1}}
\makeatother
\providecommand*\mcitethebibliography{\thebibliography}
\csname @ifundefined\endcsname{endmcitethebibliography}
  {\let\endmcitethebibliography\endthebibliography}{}

\begin{acknowledgement}

The authors thank Profs Andrew Archer, Kevin Dorfman, Tomonari Dotera, Jacob Kirkensgaard, Ron Lifshitz and An-Chang Shi for stimulating conversations, \revision{and to three anonymous referees for their constructive comments}.
MJ is grateful for a PhD studentship from the Soft Matter and Functional Interfaces (SOFI) Centre for Doctoral Training and the School of Mathematics, University of Leeds.
This work was supported by the Engineering and Physical Sciences Research Council [grant numbers EP/L015536/1, EP/P015611/1]; and the Leverhulme Trust [grant number RF-2018-449/9].
The data associated with this paper are openly available from the University of Leeds Data Repository (\url{https://doi.org/10.5518/1879})~\cite{Joseph2025data}.
This work was undertaken on ARC4, part of the High Performance Computing facilities at the University of Leeds, UK.
For the purpose of open access, the authors have applied a Creative Commons Attribution (CC BY) license to any Author Accepted Manuscript version arising from this submission.
\end{acknowledgement}

%%%%%%%%%%%%%%%%%%%%%%%%%%%%%%%%%%%%%%%%%%%%%%%%%%%%%%%%%%%%%%%%%%%%%
%% The same is true for Supporting Information, which should use the
%% suppinfo environment.
%%%%%%%%%%%%%%%%%%%%%%%%%%%%%%%%%%%%%%%%%%%%%%%%%%%%%%%%%%%%%%%%%%%%%
\begin{suppinfo}

\section{Free energy calculation of Strongly Segregated Polygons}

In this Supporting Information, we will give details of how we determine the interfacial energy and stretching free energy per chain in a Strongly Segregated Polygon (SSP).
We also give phase diagrams in the \revision{$N=300$ and $N=10,000$ cases}.

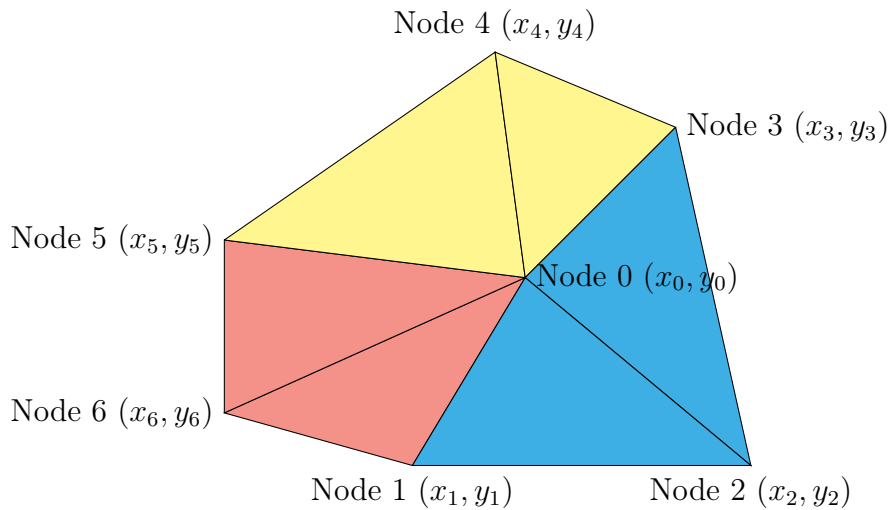
\begin{figure}
	\centering
	\begin{tikzpicture}
		\draw [fill=Salmon] (-1.5,-2.5) node[below]{Node 1 $(x_1,y_1)$}--(0,0)--(-4,-1.8)node[left]{Node 6 $(x_6,y_6)$}--(-1.5,-2.5) ;
		\draw [fill=Salmon] (-4,-1.8)--(0,0)--(-4,0.5)node[left]{Node 5 $(x_5,y_5)$}--(-4,-1.8);
		\draw [fill=Khaki1] (-4,0.5)--(0,0)--(-0.4,3)node[above]{Node 4 $(x_4,y_4)$}--(-4,0.5);
		\draw [fill=Khaki1] (0,0)--(-0.4,3)--(2,2)node[right]{Node 3 $(x_3,y_3)$}--(0,0);
		\draw [fill=CornflowerBlue] (2,2)--(0,0)--(3,-2.5)node[below]{Node 2 $(x_2,y_2)$}--(2,2);
		\draw [fill=CornflowerBlue] (0,0)--(3,-2.5)--(-1.5,-2.5)--(0,0)node[right]{Node 0  $(x_0,y_0)$};
	\end{tikzpicture}
	\caption[Structure of an \hbox{SSP}]{The structure of an \hbox{SSP} with seven nodes. All the nodes Node $i$ and their position coordinates are given. We assign red, blue and yellow colors to domains of A, B and C monomer types.}
	\label{ch3fig:SSPschematics}
\end{figure}    

The six-sided \hbox{SSP} is the basic unit in our method, and an example is given in \cref{ch3fig:SSPschematics}.
The polygon is divided into three regions as indicated by different colors in \cref{ch3fig:SSPschematics}, where each  region is assigned to different monomer types.
The color map is red for~A, blue for~B and yellow for~C.
The six nodes on the perimeter are defined by their $x$ and $y$ coordinates.
The position of the central node, which is the core of the ABC star terpolymer, follows from these positions and from the area fractions of A, B and~C, as explained below.
The central node needs to be inside the polygon for it to be valid.
Subject to this constraint, the six perimeter nodes have freedom to move around anywhere.

The SSP is divided into six triangles: $012$, $023$, $034$, $045$, $056$ and $061$, where node~0 is the central node and nodes~1 to 6 are counterclockwise (in this case) around the perimeter. 
In \cref{ch3fig:SSPschematics}, the interfaces are line segments between nodes $0$ and~$1$ (between polymers A and~B), between nodes $0$ and~$3$ (between B and~C) and between nodes~$0$ and~$5$ (between C and~A).
The contributions from these are added up to give the interfacial energy for the~\hbox{SSP}.
The stretching free energy per chain can be computed for each triangle and added up.
Thus the free energy per chain for the ABC stars within the \hbox{SSP} is determined uniquely by the positions of the nodes, and the free energy minimization can be performed by adjusting the node positions.
  
In order to determine the position of node~0, we relate the monomer compositions to the areas covered by the three types of triangle  as follows:
\begin{equation}\label{ch3eq:SSPmonomer compositions}
	\begin{aligned}
		\phi_{A}&=\frac{A_{061}+A_{056}}{A_{T}},\\
		\phi_{B}&=\frac{A_{012}+A_{023}}{A_{T}},\\
		\phi_{C}&=\frac{A_{034}+A_{045}}{A_{T}},\\
	\end{aligned}
\end{equation}
where $A_T$ is the total area of the~\hbox{SSP}. 
The area of each triangle is expressed in terms of the node coordinates in vector form.
The position vector of node~$i$ with respect to the core (node~0) is $\vec{r}_{0i}$. Following this labelling convention, the signed (vector) areas are:
\begin{equation}\label{ch3eq:Area of triangles}
	\begin{split}
		\vec{A}_{061}={}&\frac{1}{2}(\vec{r}_{06}\times \vec{r}_{01});\quad
		\vec{A}_{012}=\frac{1}{2}(\vec{r}_{01}\times\vec{r}_{02});\\
		\vec{A}_{023}={}&\frac{1}{2}(\vec{r}_{02}\times\vec{r}_{03});\quad
		\vec{A}_{034}=\frac{1}{2}(\vec{r}_{03}\times\vec{r}_{04});\\
            \vec{A}_{045}={}&\frac{1}{2}(\vec{r}_{04}\times\vec{r}_{05});\quad
		\vec{A}_{056}=\frac{1}{2}(\vec{r}_{05}\times\vec{r}_{06}).
    \end{split}
\end{equation}
We use the signs of these areas to check the validity of the polygon:
any polygon in which some areas are positive and some are negative is invalid, since this would imply that the triangles overlap.
This statement is equivalent to the requirement that the core (node~$0$) should be inside the~\hbox{SSP}.
In calculations, we use positive areas, so $A_{061}=\big|\vec{A}_{061}\big|$, and so on.
The total unsigned area~$A_T$ is the sum of the six triangle areas.

The three equations in \cref{ch3eq:SSPmonomer compositions}, with the incompressibility constraint $\phi_A+\phi_B+\phi_C=1$, can be solved for $(x_0, y_0)$ to determine the position of the core as a function of the monomer compositions and the known coordinates of other six nodes. On solving, the position of the core is
\begin{equation}\label{ch3eq:x0y0}
	\begin{split}
		x_0={}&\frac{\left(2(1-\phi_{A}-\phi_{B})A_T-x_3y_4-x_4y_5+x_4y_3+x_5y_4\right)(x_1-x_5)}{(y_3-y_5)(x_1-x_5)-(y_5-y_1)(x_5-x_3)}-{}\\
		&\quad\frac{\left(2\phi_{A}A_T+x_6y_5+x_1y_6-x_5y_6-x_6y_1\right)(x_5-x_3)}{(y_3-y_5)(x_1-x_5)-(y_5-y_1)(x_5-x_3)},\\
        y_0={}&\frac{\left(2(1-\phi_{A}-\phi_{B})A_T-x_3y_4-x_4y_5+x_4y_3+x_5y_4\right)(y_1-y_5)}{(y_3-y_5)(x_1-x_5)-(y_5-y_1)(x_5-x_3)}-{}\\
		&\quad\frac{\left(2\phi_{A}A_T+x_6y_5+x_1y_6-x_5y_6-x_6y_1\right)(y_5-y_3)}{(y_3-y_5)(x_1-x_5)-(y_5-y_1)(x_5-x_3)}.
	\end{split}
\end{equation}
The denominator in these expressions is zero only when Nodes~1, 3 and 5 are co-linear, which corresponds to the (physically impossible) situation where the internal AB, AC and BC interfaces are co-linear.
The calculation for the eight-sided SSPs is similar.

\subsection{Interfacial energy in an SSP}

In the \hbox{SSP}, the three interface lengths are $l_{01}$, $l_{03}$ and~$l_{05}$, the lengths between node~0 and nodes~1, 3 and 5, which have not (yet) been scaled by the length $R=\sqrt{Nb^2}$.
The polygon extends along the unscaled length~$d$ of the core cylinder in the third dimension.
The interfacial energy at each surface~$IJ$ (between chains of type $I$ and $J$) will be the surface tension $\tilde{\gamma}_{IJ}$ times the area of that surface,
where $\tilde{\gamma}$ is units of $k_BT$ per unit area.
The total interfacial energy, in units of~$k_BT$, is
\begin{equation}
    \label{ch3eq:interfacialssp_area}
    F_{int}=\tilde{\gamma}_{AB}\tilde{A}_{AB}+\tilde{\gamma}_{BC} \tilde{A}_{BC}+\tilde{\gamma}_{AC} \tilde{A}_{AC},
\end{equation}
where $\tilde{A}_{AB}=dl_{01}$, $\tilde{A}_{BC}=dl_{03}$ and $\tilde{A}_{AC}=dl_{05}$ are the interfacial areas.
The number of ABC star terpolymer chains in the volume is $\tilde{A}_T d/v_p$, where $v_p$ is the total volume of each terpolymer chain.
Thus, the interfacial energy per chain~$f_{int}$, in units of~$k_BT$, is
\begin{equation}\label{ch3eq:interfacial energy per chain polygon}
    f_{int}=\left(\tilde{\gamma}_{AB} l_{01}d+\tilde{\gamma}_{BC} l_{03}d+\tilde{\gamma}_{AC} l_{05}d\right)\times\frac{v_p}{\tilde{A}_T d}.
\end{equation}
We recall the relationship between the surface tension~$\tilde{\gamma}$ and the Flory interaction parameter~$\chi$:
 \begin{equation}
 \tilde{\gamma}_{IJ} = \sqrt{\frac{\chi_{IJ}}6}\rho b,
\end{equation}
where $\rho=N/v_p$ is the number of monomers per unit volume and $b$~is the step length per monomer.
We manipulate the expression for $f_{int}$ to obtain:
\begin{equation}
    f_{int}=\left(\sqrt{\frac{N\chi_{AB}}{6}} l_{01}+\sqrt{\frac{N\chi_{BC}}{6}} l_{03}+\sqrt{\frac{N\chi_{AC}}{6}} l_{05}\right) \times \frac{b\sqrt{N}}{\tilde{A}_T}.
\end{equation}
We next scale all lengths by $R=\sqrt{Nb^2}$, writing $L_{01}=l_{01}/R$ and writing $\tilde{A}_T=R^2A_T$, as in the main paper, to obtain
\begin{equation}
    f_{int}=\frac{1}{A_T}\left(\sqrt{\frac{N\chi_{AB}}{6}} L_{01}+\sqrt{\frac{N\chi_{BC}}{6}} L_{03}+\sqrt{\frac{N\chi_{AC}}{6}} L_{05}\right).
\end{equation}
We now define scaled surface tensions
\begin{equation}
  \gamma_{IJ} = \sqrt{\frac{N\chi_{IJ}}{6}}
\end{equation}
to obtain the final expression for interfacial energy per chain in units of $k_BT$ for an \hbox{SSP}:
\begin{equation}\label{ch3eq:interfacial SSP}
	f_{int}=\frac{1}{A_T}\left(\gamma_{AB}L_{01}+\gamma_{BC} L_{03}+\gamma_{AC} L_{05}\right)
\end{equation}
The interfacial energy per chain is now in terms of the position coordinates of the nodes.

\begin{figure}
	\begin{subfigure}[b]{0.45\linewidth}
		\centering
		\begin{tikzpicture}
			\draw[fill=Salmon,fill opacity=0.7] (0,0)node[left]{Node 6 }node[below]{$t=0$}--(4,0)node[right]{Node 1 }node[below]{$t=1$} --(2.45,2.67)node[above]{Node 0}--(0,0);
			\draw[dashed] (2.45,2.67)--(2.45,0)node[below]{$t_0$}node[midway,right]{$H_{061}$};
			\draw[fill=Salmon,fill opacity=1] (2.45,2.67)--(1.7,0)--(1.45,0)node[below]{$t$}--(2.45,2.67);
		\end{tikzpicture}
		\caption{}
	\end{subfigure}
	\hfill
	\begin{subfigure}[b]{0.45\linewidth}
		\centering
		\begin{tikzpicture}
			\draw[fill=Salmon,fill opacity=0.7] (0,0)node[left]{Node 6 }node[below]{$ t=0 $}--(2,0)node[right]{Node 1 }node[below]{$ t=1 $} --(3.45,2.67)node[above]{Node 0 }--(0,0);
			\draw[dashed] (3.45,2.67)--(3.45,0)node[midway,right]{$ H_{061} $}node[below]{$t_0 $}--(1.8,0) ;
			\draw[fill=Salmon,fill opacity=1] (3.45,2.67)--(1.15,0)--(1.47,0)--(3.45,2.67);
		\end{tikzpicture}
		\caption{}
	\end{subfigure}
	\caption[Possible triangles in an \hbox{SSP}]{Example of two different	possible  geometries of triangle~$061$. 
    The base length $L_{061}$ is parametrized by~$t$.
    The height~$H_{061}$ of the triangle is marked in both cases, and it intersects the base at~$t_0$.
    The point of intersection can be within the triangle, and in~(a), or outside the triangle, as in~(b).}
	\label{fig:differernt_triangles_for_A}
\end{figure}
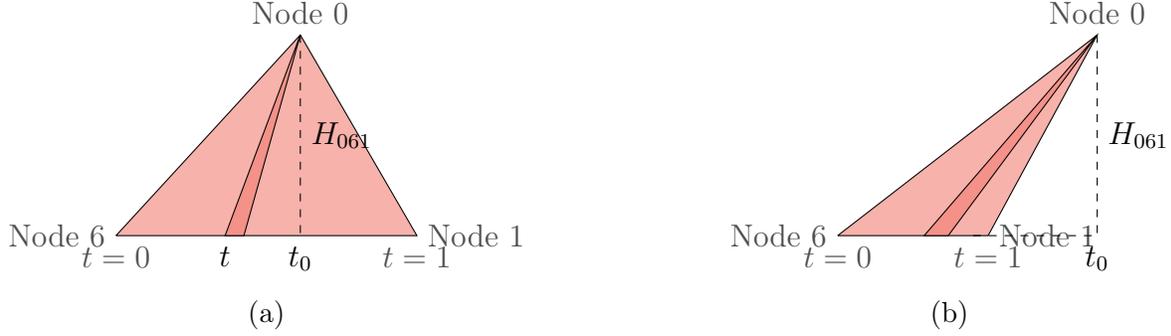

\subsection{Stretching energy in an SSP}

In this section, we work directly in scaled units, with lengths written in units of $R=\sqrt{Nb^2}$, and energies written in units of~$k_BT$.
The stretching free energy is determined for each triangle separately and then added up to find the total stretching energy per chain in an~\hbox{SSP}.
Node~$0$ is the core. The presence of a core introduces a logarithmic correction to the stretching free energy.
The calculation of the stretching free energy per chain is demonstrated by considering triangle~$061$, between Nodes $1$, $6$, and~$0$, of monomer type~A.

The triangle~$061$ can have any triangular configuration.
The perpendicular height of the triangle can intersect the base either inside or outside the triangle, as illustrated in \cref{fig:differernt_triangles_for_A}. 
The location of the wedge along the base of the triangle is parameterized by a variable~$t$, which can take values from $0$ to~$1$, between nodes~$6$ ($t=0$) and~$1$ ($t=1$).
To determine the stretching free energy of the triangle, we calculate the stretching free energy for the wedge and sum over the length~$L_{061}$.

In vector notation the nodes are written as $\vec{r}_0=[x_0,y_0]^T$, $\vec{r}_1=[x_1,y_1]^T$ and $\vec{r}_6=[x_6,y_6]^T $. 
The perpendicular from node~$0$ intersects the base at~$\vec{r}_h$, which is at a point parameterized by~$t_0$, so $\vec{r}_h=\vec{r}_6+t_0\left(\vec{r}_1-\vec{r}_6\right)$.
Since $\left(\vec{r}_h-\vec{r}_0\right)$ is perpendicular to $\left(\vec{r}_6-\vec{r}_1\right)$, the dot product between these is zero, which gives~$t_0$:
\[
t_0=\frac{\left(\vec{r}_6-\vec{r}_0\right)
          \cdot
          \left(\vec{r}_6-\vec{r}_1\right)}
          {\left|\vec{r}_6-\vec{r}_1\right|^2}.
\]
Now that $t_0$, and hence~$\vec{r}_h$, is obtained in terms of the positions of nodes~0, 1 and~6, the height of the triangle is $H_{061}=|\vec{r}_h-\vec{r}_0|$ and the length of the base of the triangle is $L_{061}=|\vec{r}_6-\vec{r}_1|$.
With $t_0$ as given above, the height of a wedge~$H(t)$ at any $t$ is then 
 \begin{equation}
     \label{ch3eq:Rwfor wedge inSSP}
     H(t)=\sqrt{((t-t_0)L_{061})^2 +H_{061}^2}.
 \end{equation}

We now consider the triangle in \cref{fig:differernt_triangles_for_A}(a) to be extended in the third dimension to a depth~$D$ (in scaled length units), giving a volume $A_{061}D$.
The SSP, with area~$A_T$, has corresponding volume~$A_T D$. 
A single ABC star terpolymer occupies volume~$V_p$ (in scaled units), so the total number of ABC star terpolymers in the SSP is $A_T D/V_p$.
An A-block occupies volume~$\phi_A V_p$, so the total number of A-blocks in the triangle is $A_{061}D/(\phi_A V_p)$.
The number of A-blocks in the wedge spanned by increment~$dt$ is $A_{061}D/(\phi_A V_p) \times dt$, since the area of a wedge of width~$dt$ is $A_{061}dt$, independent of the position of the wedge.
The stretching energy per A-block in a wedge, as given in \cref{ch3eq:Stretching_with_convex}, is
\begin{equation}\label{ch3eq:Stretching_with_convex_supp}
   f_{chain}(H(t),\phi_A) = \frac{3}{4\phi_A}H(t)^2\log(cH(t)^2) = \frac{3}{4\phi_A}F_{chain}(H(t)),
\end{equation}
where $c=R^2/R_{core}^2$ and $F_{chain}(H)=H^2\log(cH^2)$.
Multiplying by the number of A~chains in a wedge and integrating over all wedges in the triangle gives the total free energy for triangle~061:
\begin{equation*}
    \frac{A_{061}D}{\phi_A V_p} \int_0^1 f_{chain}(H(t),\phi_A) dt.
\end{equation*}
The contribution from triangle 061 to the stretching energy per chain in the SSP can be obtained by dividing this by the total number of polymers in the SSP, giving:
\begin{equation}
    f_{061} = \frac{V_p}{A_T D}\frac{A_{061}D}{\phi_A V_p} \int_0^1 f_{chain}(H(t),\phi_A) dt = \frac{A_{061}}{\phi_A A_T} \int_0^1 f_{chain}(H(t),\phi_A) dt.
\end{equation}
We substitute from \cref{ch3eq:Stretching_with_convex_supp} using the wedge height in \cref{ch3eq:Rwfor wedge inSSP} to give
\begin{equation}\label{ch3eq:stretching_free}
	f_{061}=\frac{3A_{061}}{ 4\phi_{A}^2 A_T}\int_{0}^{1}F_{chain}\left(\sqrt{((t-t_0)L_{061})^2 +H_{061}^2} \right)dt,
\end{equation}
where $F_{chain}(H)=H^2\log(cH^2)$.

The integral above is of the general form, 
\begin{equation}\label{108}
	I_{061}(X,Y,t_0)=
    \int_{0}^{1}(X(t-t_0)^2+Y)\log(c(X(t-t_0)^2+Y))dt,
\end{equation}
where $X=L_{061}^2$, $Y=H_{061}^2$ and $H(t)=X(t-t_0)^2+Y$.
This integral can be evaluated explicitly:  
\begin{equation}\label{eq:Ifor061_supp}
	\begin{aligned}
	I_{061}=&\log(c(X(1-t_0)^2+Y))\left(\frac{X(1-t_0)^3}{3}+Y(1-t_0)\right)+{}\\
	&\log(c(Xt_0^2+Y))\left(\frac{Xt_0^3}{3}+Yt_0\right)-\frac{2X}{9}\left((1-t_0)^3+t_0^{3}\right)-\frac{4Y}{3} + {}\\
	&\frac{4Y}{3}\sqrt{\frac{Y}{X}}\left(\arctan\sqrt{\frac{X}{Y}}(1-t_0)+\arctan\sqrt{\frac{X}{Y}}t_0\right).
	\end{aligned}
\end{equation}
Hence, for triangle $061$, the stretching free energy per chain in units of $k_BT$ is
\begin{equation}\label{ch3eq:F061}
	f_{061}=\frac{3}{4}\frac{A_{061}}{\phi_{A}^2A_T}I_{061}(L_{061}^{2},H_{061}^{2},t_0),
\end{equation}
where $t_0$ is given above in terms of the positions of nodes 0, 6 and~1.

The total stretching free energy per chain~$f_{str}$ of an \hbox{SSP} is given by summing the equivalent expressions for all six triangles, recalling that
triangles 061 and 056 have monomer~A, 
triangles 023 and 012 have monomer~B, 
triangles 045 and 034 have monomer~\hbox{C}, 
and that $t_0$ will be different for each triangle.
The outcome is:
\begin{equation}\label{ch3eq:SSPstrechfree_supp}
   f_{str} = f_{061}+f_{056}+f_{045}+f_{034}+f_{023}+f_{012}.
\end{equation}

\subsection{\revision{Core contributions to free energy}}

Since we expect the core region to be of order monomer dimensions, the exact details must depend on the specific chemistry, both of the monomer type in each of the three arms, and of the chemical unit used to form the branch point itself. It is not possible to develop a ``universal'' theory.  However, we can make reasonable assumptions to arrive at a plausible first order description of the core. We assume that the chains in the core region are sufficiently closely packed together so that the arms are forced to exit the core region as quickly as possible, i.e., they are strongly stretched away from the core at the monomer scale. Thus, if the core region contains $N_{core} \ll N$ monomers per chain, we expect $R_{core} \approx N_{core}b/3$ for a three arm star. 

Now, consider an SSP of area $\tilde{A}_T=R^2A_T$ (scaled by length $R=\sqrt{Nb^2}$) and depth $d$ in the third dimension, and so of volume $\tilde{A}_T d$.  The volume per chain is $N v_0$, where $v_0$ is the volume of a single monomer unit.
Then the number of chains in the SSP is,
\begin{equation*}
\frac{\tilde{A_T}d}{Nv_0}.
\end{equation*}
But, the core region contains $N_{core}$ monomers per chain, each of volume $v_0$, so the core volume must be:
\begin{equation*}
\frac{\tilde{A_T}d}{Nv_0} N_{core} v_0 = \frac{\tilde{A_T}dN_{core}}{N}.
\end{equation*}
But the core volume is also $\pi R_{core}^2 d$, so we find:
\begin{align*}
   \pi R_{core}^2 d &=  \frac{\tilde{A_T}dN_{core}}{N}, \\
     R_{core}^2  &=  \frac{\tilde{A_T}N_{core}}{\pi N}.
\end{align*}
But we also have $R_{core} \approx N_{core}b/3$, and so:
\begin{align*}
    \frac{N_{core}^2 b^2}{9} & = \frac{\tilde{A_T}N_{core}}{\pi N}, \\
    N_{core} &= \frac{9\tilde{A_T}}{\pi Nb^2} = \frac{9A_T}{\pi}
\end{align*}
and hence:
\begin{equation*}
    R_{core} = \frac{3A_T b}{\pi}.
\end{equation*}
As $A_T$ increases, the number of chains per unit length of the core also increases, so that the radius of the core (in which chains are stretched to monomer level by chain packing) must increase. Hence, we find:
\begin{equation}
c = \frac{R^2}{R_{core}^2} \approx \frac{\pi^2 N}{9A_{T}^2},
\end{equation}
i.e., the value of $c$ used in the equations for the stretching free energy is not constant but varies with the SSP area, and depends on the degree of polymerization~$N$.

The chains inside the core region are strongly stretched to the monomer level, and we estimate the energy for this to be of order $k_B T$ per monomer, giving an energy per chain (in units of $k_B T$) of:
\begin{equation*}
    f_{st,core} = s_{core} N_{core}  = s_{core} \frac{9A_T}{\pi},
\end{equation*}
where $s_{core}$ is a parameter expected to be order one.  

Monomers within the core are also necessarily brought into close proximity, and we expect the number of monomers of each type (A, B or C) in the core to be approximately equal.  Hecne, we assume the composition of the core is approximately $\phi_A =\phi_B = \phi_C=\frac{1}{3}$, giving an energy per chain of:
\begin{align*}
    f_{int,core} &= N_{core} (\chi_{AB}\phi_A\phi_B+\chi_{BC}\phi_B\phi_C+\chi_{AC}\phi_A\phi_C) \\  
      &=  \frac{9A_T}{\pi} \left(  \frac{\chi_{AB}+\chi_{BC}+\chi_{AC}}{9} \right). 
\end{align*}
Adding these together gives our proposed core energy, of:
\begin{equation} \label{eq:core_energy}
   f_{core} = f_{st,core} + f_{int,core} = \frac{9A_T}{\pi} \left( s_{core} + \frac{\chi_{AB}+\chi_{BC}+\chi_{AC}}{9} \right). 
\end{equation}

We add the stretching free energy~\cref{ch3eq:SSPstrechfree_supp} to the interfacial free energy~\cref{ch3eq:interfacial SSP} and core energy~\cref{eq:core_energy} to obtain the total free energy per chain of an \hbox{SSP}~$f_c$, in units of $k_B T$,
\begin{equation}\label{ch3eq:freeenergyf_cSSP_supp}
 f_c=f_{int}+f_{str}+f_{core}.
\end{equation}
This total free energy is now given explicitly in terms of the positions of the six nodes, the monomer compositions~$(\phi_A,\phi_B,\phi_C)$ (with $\phi_A+\phi_B+\phi_C=1$), the Flory interaction parameters and the degree of polymerization.

\subsection{Phase diagrams with different values of $N$}

\begin{figure}
    \centering
\includegraphics[width=1.0\linewidth]{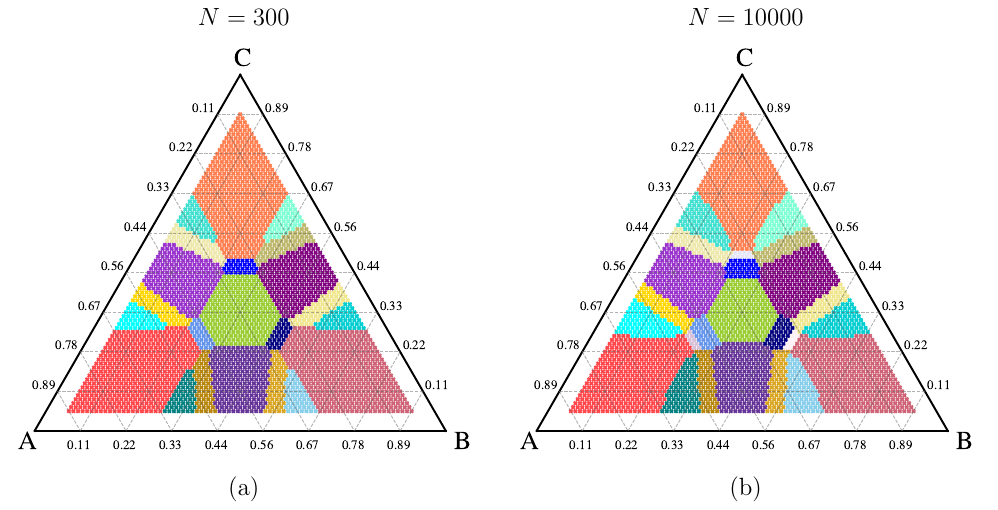}
\caption{Phase diagrams with different values of $N$: (a)~$N=300$; (b)~$N=10000$. Compare with diagram for $N=1000$, shown in \cref{fig:symmetric_phase_diagram}(b). 
The data for this figure is available from~\cite{Joseph2025data}.}
    \label{fig:largerc}
\end{figure}

\Cref{fig:largerc} shows phase diagrams with $N=300$ and $N=10000$. 
Compared to $N=1000$ (\cref{fig:symmetric_phase_diagram}b), the overall placement of regions with stable morphologies is  qualitatively unchanged, though there are subtle changes in the areas covered by each morphology. 
\revision{The $[10.6.4;10.4.6;10.6.6]$ morphology, masked by L+C in \cref{fig:symmetric_phase_diagram}(b), appears in phase diagram corresponding to higher degree of polymerization $N = 10,000$ in \cref{fig:largerc}.(b), with two rows of $[10.6.4;10.4.6;10.6.6]$ (light blue regions next to orange lamellar L+C regions).}

\end{suppinfo}

%%%%%%%%%%%%%%%%%%%%%%%%%%%%%%%%%%%%%%%%%%%%%%%%%%%%%%%%%%%%%%%%%%%%%
%% The appropriate \bibliography command should be placed here.
%% Notice that the class file automatically sets \bibliographystyle
%% and also names the section correctly.
%%%%%%%%%%%%%%%%%%%%%%%%%%%%%%%%%%%%%%%%%%%%%%%%%%%%%%%%%%%%%%%%%%%%%

\end{document}